\documentclass[fleqn,usenatbib]{mnras}

\usepackage{newtxtext,newtxmath}

\usepackage[T1]{fontenc}
\usepackage{ae,aecompl}

\usepackage{graphicx}	
\usepackage{amsmath}	
\usepackage{amssymb}	

\newcommand{\beqa}{\begin{eqnarray}} 
\newcommand{\eeqa}{\end{eqnarray}}

\newcommand{\bsub}{\begin{subequations}}
\newcommand{\esub}{\end{subequations}}
\newcommand{\beal}{\begin{align}}
\newcommand{\ealn}{\end{align}}

\title[Nickel and iron in SNe Ia]{
Sub-Chandrasekhar progenitors favoured for type Ia supernovae: Evidence from late-time spectroscopy\thanks{Based on observations collected with ESO telescopes under programme IDs 086.D-0747, 087.D-0161, 088.D-0184, 090.D-0045, 095.A-0316, 096.D-0829, 097.D-0967, 098.D-0692 and 0100.D-0285}.
}
\author[A. Fl\"ors et al.]{A.~Fl\"ors$^{1,2,3}$\thanks{E-mail: afloers@mpa-garching.mpg.de}, J.~Spyromilio$^{1}$, S.~Taubenberger$^{2}$, S.~Blondin$^{4}$, R.~Cartier$^{5}$, 
\newauthor B.~Leibundgut$^{1}$, L.~Dessart$^{4}$, S.~Dhawan$^{6}$ and W.~Hillebrandt$^{2}$
\\
$^{1}$European Southern Observatory, Karl-Schwarzschild-Stra\ss e 2,
D-85748 Garching bei M\"unchen, Germany\\
$^{2}$Max-Planck-Institut f\"ur Astrophysik, Karl-Schwarzschild-Stra\ss e 1, D-85748 Garching bei M\"unchen, Germany\\
$^{3}$Physik-Department, Technische Universit\"at M\"unchen, James-Franck-Stra\ss e 1, D-85748 Garching bei M\"unchen, Germany\\
$^{4}$Unidad Mixta Internacional Franco-Chilena de Astronom\'ia, CNRS/INSU UMI 3386 \\~~and Instituto de Astrof\'isica, Pontificia
     Universidad Cat\'olica de Chile, Santiago, Chile\\
$^{5}$Cerro Tololo Inter-American Observatory, National Optical Astronomy Observatory, Casilla 603, La Serena, Chile\\
$^{6}$The Oskar Klein Centre, Physics Department, Stockholm University, Albanova University Center, SE 106 91 Stockholm, Sweden
}

\date{Accepted 2019 September 24. Received 2019 September 23; in original form 2019 August 16.}

\pubyear{2019}

\begin{document}
\label{firstpage}
\pagerange{\pageref{firstpage}--\pageref{lastpage}}
\maketitle

\begin{abstract}
A non-local-thermodynamic-equilibrium (NLTE) level population model of the first and second ionisation stages of iron, nickel and cobalt is used to fit a sample of XShooter optical + near-infrared (NIR) spectra of Type Ia supernovae (SNe Ia). From the ratio of the NIR lines to the optical lines limits can be placed on the temperature and density of the emission region. We find a similar evolution of these parameters across our sample. Using the evolution of the \ion{Fe}{II} 12\,570\,\AA\,to 7\,155\,\AA\,line as a prior in fits of spectra covering only the optical wavelengths we show that the 7200\,\AA\,feature is fully explained by [\ion{Fe}{II}] and [\ion{Ni}{II}] alone. This approach allows us to determine the abundance of \ion{Ni}{II}/\ion{Fe}{II} for a large sample of 130 optical spectra of 58 SNe Ia with uncertainties small enough to distinguish between Chandrasekhar mass (M$_{\text{Ch}}$) and sub-Chandrasekhar mass (sub-M$_{\text{Ch}}$) explosion models. We conclude that the majority (85\%) of normal SNe Ia have a Ni/Fe abundance that is in agreement with predictions of sub-M$_{\text{Ch}}$ explosion simulations of $\sim Z_\odot$ progenitors. Only a small fraction (11\%) of objects in the sample have a Ni/Fe abundance in agreement with M$_{\text{Ch}}$ explosion models.
\end{abstract}




\section{Introduction}
Type Ia supernovae (SNe Ia) are a remarkably homogeneous class of transients which are thought to originate from the explosion of a white dwarf (WD) star in a binary system. Radioactive \ensuremath{^{56}{\rm Ni}} produced in the thermonuclear explosion of the electron-degenerate matter \citep{1960ApJ...132..565H} powers the light curve \citep{1962PhDT........25P, 1969ApJ...157..623C, 1994ApJ...426L..89K} for several years. Even though SNe Ia have been used as distance indicators for several decades and significantly contributed to our current understanding of cosmology \citep[$\Lambda$CDM and the accelerated expansion of the universe, ][]{1998AJ....116.1009R, 1999ApJ...517..565P}, the precise mechanism that leads to the thermonuclear runaway reactions, as well as the progenitor system, remains elusive. 

Two channels that can lead to the explosion of a WD as a SNe Ia have been extensively discussed in the literature. In Chandrasekhar-mass (M$_{\text{Ch}}$) explosions the burning front propagates either as a deflagration \citep[e.g.][]{2003Sci...299...77G, 2014MNRAS.438.1762F} or a delayed detonation \citep[e.g.][]{1986SvAL...12..131B, 1987SvAL...13..364B, 1991A&A...245..114K, 2005ApJ...623..337G, 2013MNRAS.429.1156S}. The explosion is naturally triggered by an increase of the central density as the WD accretes material from its companion and comes close to the Chandrasekhar mass limit ($M\simeq\,$M$_{\text{Ch}}$). In the sub-M$_{\text{Ch}}$ channel the central temperature of the primary white dwarf never reaches conditions that are sufficient to ignite carbon. However, an explosion significantly below the M$_{\text{Ch}}$ may be triggered through dynamical processes such as mergers \citep[e.g.][]{2010Natur.463...61P, 2013ApJ...770L...8P, 2013MNRAS.429.1425R}, double detonations \citep[e.g.][]{2010A&A...514A..53F, 2011ApJ...734...38W, 2013ApJ...774..137M, 2018ApJ...854...52S} or head-on collisions \citep[e.g.][]{2013ApJ...778L..37K}. For such systems the burning front propagates as a pure detonation \citep{2010ApJ...714L..52S, 2018MNRAS.474.3931B}.

The search for solutions to the SN Ia progenitor problem has been the focus of many studies. 
For historical supernova remnants one can search for a surviving companion star which was ejected at velocities of a few hundred km\,s$^{-1}$, though no promising candidates have been found so far (\citeauthor{2018MNRAS.479..192K} \citeyear{2018MNRAS.479..192K} for SN\,1006, \citeauthor{2018MNRAS.479.5696K} \citeyear{2018MNRAS.479.5696K} for SN\,1572). Non-degenerate donor stars of SNe Ia in nearby galaxies should also be visible in deep images as their brightness increases by a factor of $\sim10 - 10^3$, though again, no donor stars have been found for a sample of the closest SNe Ia in recent times \citep{2011Natur.480..348L, 2012ApJ...744L..17B, 2013ApJ...765..150S}. 

The growth of a white dwarf star to the M$_{\text{Ch}}$ limit requires a steady transfer of material from the companion \citep{1984ApJ...286..644N}. Material which was expelled from the companion but not accreted on the white dwarf enriches the circumstellar material (CSM). In few cases evidence for such a CSM
has been detected (\citeauthor{2003Natur.424..651H} \citeyear{2003Natur.424..651H}, \citeauthor{2004ApJ...605L..37D} \citeyear{2004ApJ...605L..37D} for SN\,2002ic; \citeauthor{2018ApJ...868...21H} \citeyear{2018ApJ...868...21H}, \citeauthor{2019ApJ...871...62G} \citeyear{2019ApJ...871...62G} for SN\,2015cp; \citeauthor{2019MNRAS.487.2372V} \citeyear{2019MNRAS.487.2372V}, \citeauthor{2019MNRAS.486.3041K} \citeyear{2019MNRAS.486.3041K} for SN\,2018fhw). However, the bulk of Ias do not exhibit any evidence for CSM interaction. When the blast wave from the SN explosion runs through the CSM, electrons are accelerated to relativistic speeds and produce radio emission through synchrotron radiation \citep{1982ApJ...259..302C, 1998ApJ...499..810C, 2006ApJ...651..381C}. For a nearby SN Ia such as SN\,2011fe radio emission should be observable if it exploded in the single degenerate channel. However, no radio emission was found by \citet{2012ApJ...746...21H} for SN\,2011fe (but see also \citet{2011Natur.480..344N} for a counter argument)

One can also distinguish the two channels from direct observations of the aftermath of the explosion itself. Unfortunately, the uniformity of explosion model predictions of SNe Ia makes this a challenging task. A promising difference between M$_{\text{Ch}}$ and sub-M$_{\text{Ch}}$ models is the mass fraction of neutronized species produced in the explosion. While the progenitor metallicity affects how much neutron-rich material can be produced in both channels, additional neutrons are only available for explosions close to the M$_{\text{Ch}}$ due to the high central densities ($\rho_{\text{cen}} \sim 2\times 10^9$\,g\,cm$^{-3}$) which allow electron capture reactions to take place \citep{1999ApJS..125..439I, 2013MNRAS.429.1156S}. 

X-ray spectroscopy of SN remnants in the Milky Way (MW) and the Large and Small Magellanic Clouds (LMC \& SMC) allowed \citet{2013ApJ...767L..10P}, \citet{2014HEAD...1430402Y}, \citet{2015ApJ...801L..31Y}, \citet{2017ApJ...843...35M} and \citet{2019arXiv190605972S} to estimate the fraction of the neutron-rich stable iron-peak isotopes $^{55}$Mn and $^{58}$Ni. They find considerable differences across their sample, but the number of objects for which such a study can be done is limited. 

In this work we are interested in the composition of the iron-rich ejecta of SNe Ia. Theoretical explosion models contain the following isotopes in the central region:
\begin{itemize}
    \item[a)] $^{56}$Ni, which is the most abundant radioactive isotope and responsible for the heating of the ejecta. It decays within a few days ($t_{1/2}=6.075\,$d) to $^{56}$Co, which in turn decays ($t_{1/2}=77.2\,$d) to stable $^{56}$Fe. $^{56}$Ni can be produced in NSE (Nuclear Statistical Equilibrium) without an overabundance of neutrons (Y$_e=0.5$) or high densities \citep{1960ApJ...132..565H}. In our analysis, we treat $^{56}$Ni as a reference point and give other abundances in fractions of the $^{56}$Ni mass.
    \item[b)] $^{57}$Ni, which decays with $t_{1/2}=1.48\,$d to $^{57}$Co. The decay of $^{57}$Co to stable $^{57}$Fe is slower ($t_{1/2}=271.74\,$d) than the decay of $^{56}$Co, so it can power the light curve at later epochs. Roughly 1\,000\,days after the explosion energy deposition from $^{57}$Ni decay overtakes the energy deposition from $^{56}$Ni \citep{2009MNRAS.400..531S}. Most sub-M$_{\text{Ch}}$ explosions models predict an abundance M$_{^{57}\text{Ni}}$/M$_{^{56}\text{Ni}}$\,<\,2\,\% \citep[e.g.][]{2010ApJ...714L..52S, 2010Natur.463...61P, 2015ApJ...801L..31Y, 2018SSRv..214...67N, 2018ApJ...854...52S}, while M$_{\text{Ch}}$ explosions predict >\,2\,\% \citep[e.g.][]{2013MNRAS.429.1156S, 2015ApJ...801L..31Y, 2018SSRv..214...67N}.
    \item[c)] stable $^{54,56}$Fe which is directly synthesized in the explosion and not a daughter product of radioactive decay. M$_{\text{Ch}}$ explosions produce M$_{^{54,56}\text{Fe}}$/M$_{^{56}\text{Ni}}$\,>\,10\,\%, while most sub-M$_{\text{Ch}}$ models have M$_{^{54,56}\text{Fe}}$/M$_{^{56}\text{Ni}}$\,<\,10\,\% (see references in b).
    \item[d)] stable $^{58}$Ni which is synthesized in the explosion. Sub-M$_{\text{Ch}}$ explosions contain M$_{^{58}\text{Ni}}$/M$_{^{56}\text{Ni}}$\,<\,6\,\% and M$_{\text{Ch}}$ explosions have M$_{^{58}\text{Ni}}$/M$_{^{56}\text{Ni}}$ between 8 and 12\,\% (see references in b). 
\end{itemize}

Contributions of slowly-decaying neutron-rich material (e.g. $^{57}$Co - the daughter product of $^{57}$Ni) to the quasi-bolometric light curve of the nearby SN\,2011fe at >1000\,days after the explosion were investigated by \citet{2017hst..prop15192S}, \citet{2017MNRAS.468.3798D} and \citet{2017MNRAS.472.2534K}. This method was used for other nearby transients SN\,2012cg \citep{2016ApJ...819...31G}, SN\,2013aa \citep{2018ApJ...857...88J}, SN\,2014J \citep{2018ApJ...852...89Y}, SN\,2014lp \citep{2018ApJ...866...10G} and SN\,2015F \citep{2018ApJ...859...79G}. However, the physical processes relevant at such late phases (e.g. ionization/recombination) have long time constants and their onset is poorly constrained by the data \citep{2015ApJ...814L...2F}. In particular, it remains unclear what fraction of the radioactive decay energy is converted into optical photons as the majority of the energy is expected to come out in the mid-IR. 

SNe Ia complete their transition into the nebular phase roughly half a year after the explosion when the ejecta become fully transparent to optical and NIR photons and the bare iron core which gives insight into the explosion physics is visible. Nebular phase spectral models build on the early work of \citet{1980PhDT.........1A} and many authors over the years \citep{1992MNRAS.258P..53S, 1992ApJ...390..602K, 1994ApJ...426L..89K, 2005A&A...437..983K, 2007Sci...315..825M, 2015ApJ...814L...2F, 2017ApJ...845..176B, 2018MNRAS.477.3567M, 2018ApJ...861..119D} with spectral synthesis codes of varying complexity.

The method presented herein enables the use of optically thin Ni and Fe lines at optical and NIR wavelengths to constrain the fraction of neutron rich material in the ejecta. This approach increases the number of objects for which the analysis can be performed by about one order of magnitude compared to the number of optical+NIR spectra currently available. The analysis is made possible by using a small sample of optical+NIR spectra to determine the relative strength of the NIR 12\,570\,\AA\,to the optical emission 7\,155\,\AA\,lines of \ion{Fe}{II}. With this relation we can model optical nebular spectra which do not have NIR observations. From the fits to observed late spectra we determine the stable Ni to Fe ratio. We estimate the systematic uncertainty of the method and show that emission lines from singly ionized iron and nickel are sufficient to model the 7\,200\,\AA\,emission feature. Finally, we discuss the implications of the determined Ni to Fe ratio of the large sample of more than 100 spectra on the various explosion model predictions.
\section{Observations}
\label{SectionObservations}
We extend the XShooter sample of nearby SNe Ia in the nebular phase from \citet{2018MNRAS.477.3567M} with SN\,2015F (PI M. Sullivan, program ID 095.A-0316, PI R. Cartier, program IDs 096.D-0829, 097.D-0967 and 098.D-0692) and SN\,2017bzc (PI L. Dessart, program ID 0100.D-0285). The epochs of the additional spectra range from $\sim$\,200 days to $\sim$\,420 days after the explosion. An overview of the observations for these two supernovae is given in Table~\ref{tab:Observations}.

XShooter is an echelle spectrograph with three arms (UVB, VIS and NIR) covering the wavelength range of $\sim$\,3\,000--25\,000\,\AA\,located at the Very Large Telescope (VLT). The resolution of the individual arms depends on the slit widths. For the observations presented in this work slit widths of 1.0 (UVB), 0.9 (VIS) and 0.9 (NIR) arcseconds have been used. The corresponding resolution of the three arms is therefore 5400, 8900 and 5600, respectively. The spectra were reduced using the ESO pipeline with the XShooter module, producing flux-calibrated one-dimensional spectra in each of the three arms \citep{2010SPIE.7737E..28M,2013A&A...559A..96F}. We also used a custom postprocessing pipeline to combine the rectified 2D-images, perform the sky-subtraction and extract the spectrum (\url{https://github.com/jselsing/xsh-postproc}).

\begin{table*}
	\caption{Overview of the observations}
	\label{tab:Observations}
    \begin{tabular}{lcclccccc}
    	\hline 
    	SN name   & Observation & Observation & Phase$^a$     & E($B-V$)$^b$ & Helio. z$^c$ & Host galaxy & Exposure \\
                  & MJD         & date        &  & (mag)    &          &             & time (s) \\
		\hline                   
        SN\,2015F   & 57287.4     & 2015 Sept 22& +181d & 0.260$\pm$0.021$^d$ & 0.00489 & NGC 2442 & 720\\
                  & 57331.4     & 2015 Nov 05 & +225d &                     & & & 1200\\
                  & 57345.3     & 2015 Nov 19 & +239d &                     & & & 3600\\
                  & 57372.2     & 2015 Dec 16 & +266d &                     & & & 7600\\
                  & 57512.0     & 2016 May 04 & +406d &                     & & & 3800\\
        SN\,2017bzc & 58039.5     & 2017 Oct 12 & +215d & 0.0122$\pm$0.0002   & 0.00536 & NGC 7552 & 10080\\
		\hline
			\end{tabular}
		\begin{flushleft}
         $^a$Phase of late-time spectrum calculated with respect to maximum light. \\
         $^b$Galactic $E$(\textit{B-V}) values from \protect \cite{2011ApJ...737..103S}.\\
         $^c$Heliocentric redshifts are from the Nasa Extragalactic Database (NED).\\
         $^d$Additional host galaxy extinction of $E$(\textit{B-V})=0.085 mag was found for SN 2015F by \citet{2017MNRAS.464.4476C}. The value given in the table is the combined MW and host galaxy $E$(\textit{B-V}). \\
 \end{flushleft}
 \end{table*}

\begin{figure*}
    \centering
    \begin{minipage}{.475\textwidth}
        \centering
        \includegraphics[width=1\linewidth]{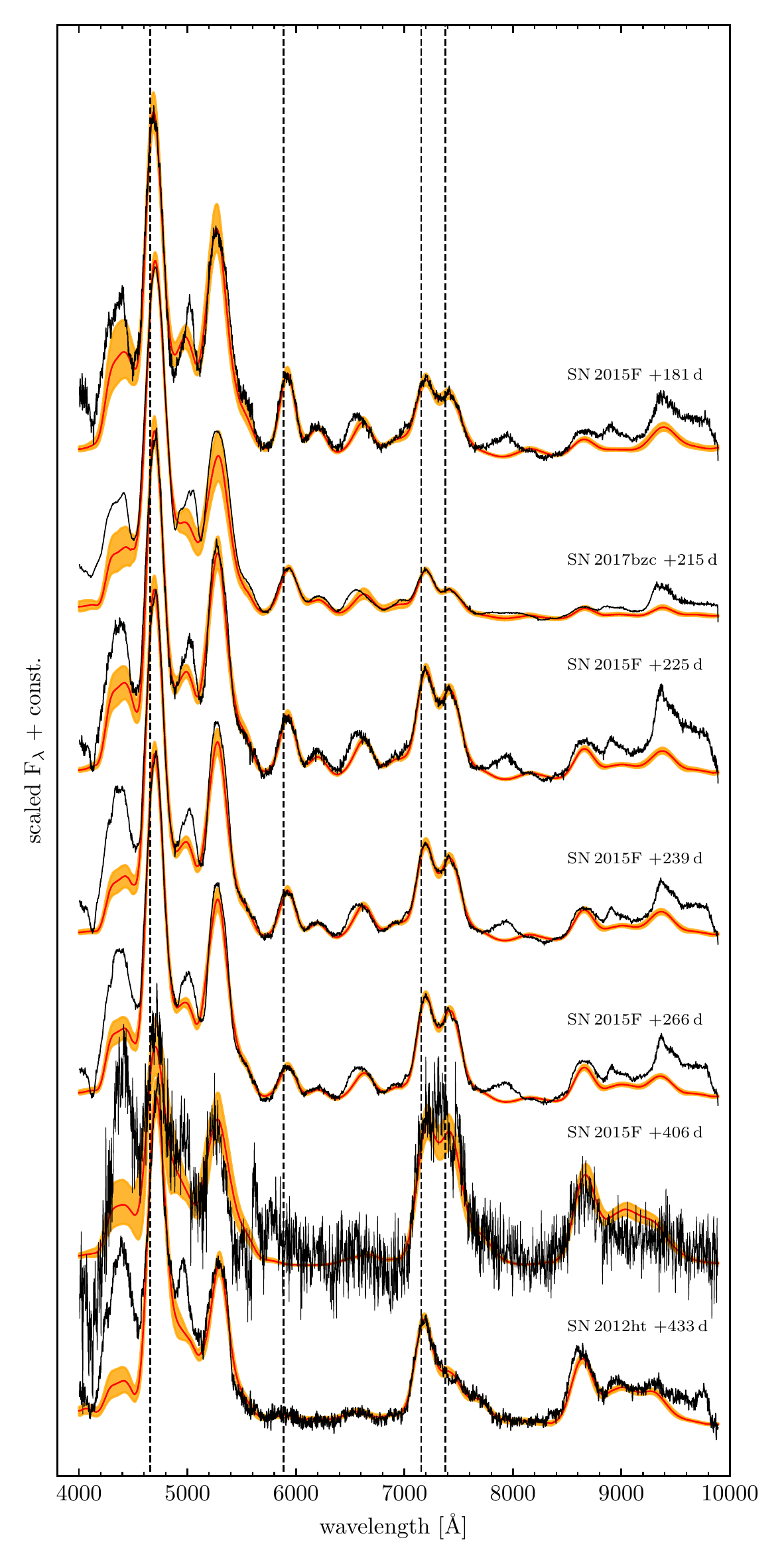}
    \end{minipage}%
    \begin{minipage}{.475\textwidth}
        \centering
        \includegraphics[width=1\linewidth]{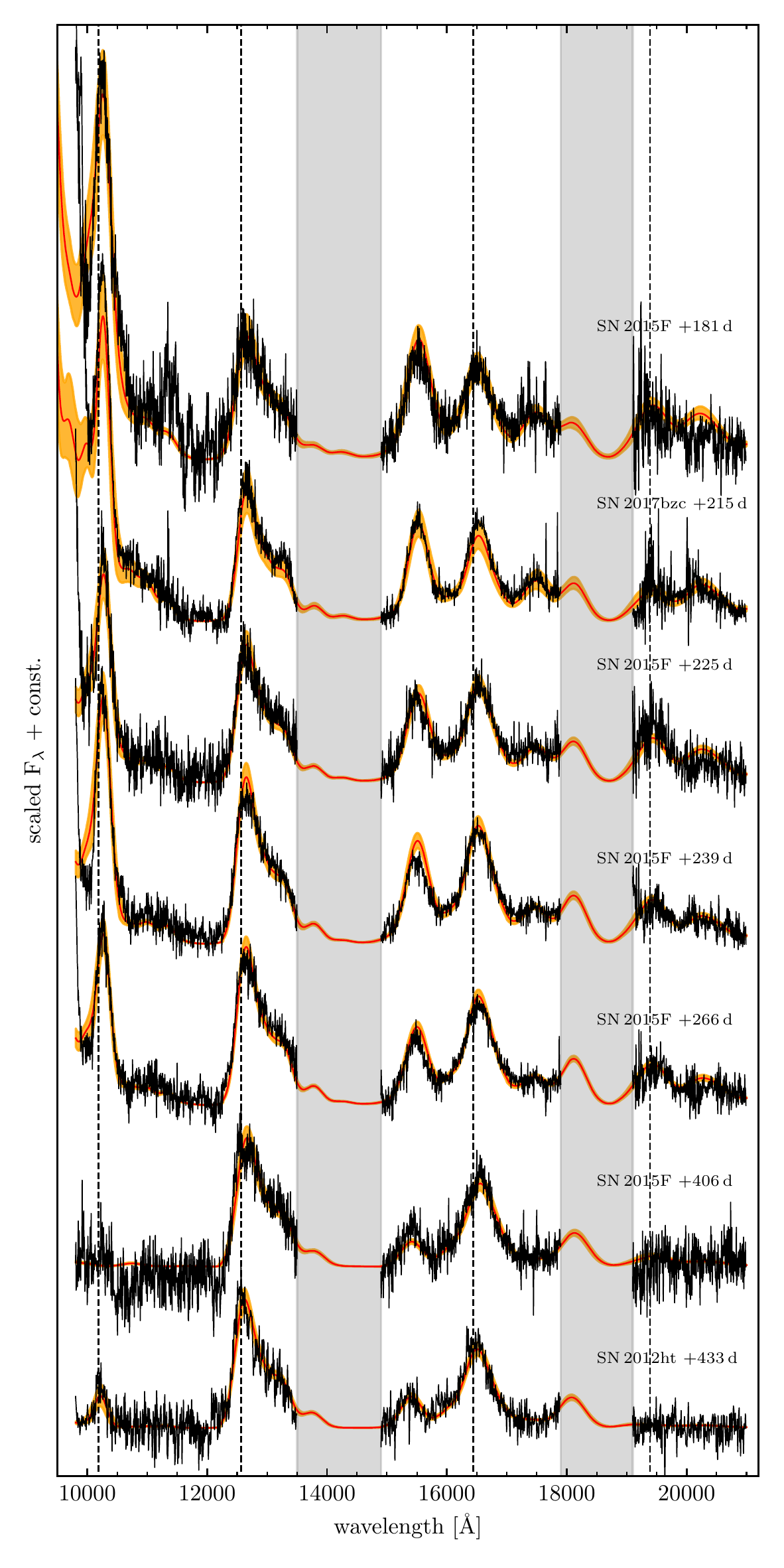}
    \end{minipage}
    \caption{Optical (left) and NIR (right) spectra of SN\,2015F and SN\,2017bzc obtained with XShooter at the VLT. We also show the spectrum and the corresponding fit to SN\,2012ht presented in \citet{2018MNRAS.477.3567M}. The spectra are arranged in epoch starting with the youngest at the top. The spectra have been corrected for telluric absorption but not for extinction and host galaxy redshift. Instead, we redshift and extinguish the spectral models. Fluxes are normalized to the $4\,700\,$\AA\,[\ion{Fe}{III}] feature (optical) and $12\,600\,$\AA\,[\ion{Fe}{II}] feature (NIR). In the NIR the bands of heavy telluric absorption are masked in grey. In the optical the rest wavelengths of the 4\,659\,\AA\,[\ion{Fe}{III}], the 5\,888\,\AA\,[\ion{Co}{III}], the 7\,155\,\AA\,[\ion{Fe}{II}] and the 7\,378\,\AA\,[\ion{Ni}{II}] lines are indicated as dashed lines. In the NIR dashed lines indicate the 10\,190\,\AA\,[\ion{Co}{II}], the 12\,570\,\AA\,[\ion{Fe}{II}], the 16\,440\,\AA\,[\ion{Fe}{II}] and the 19\,390\,\AA\,[\ion{Ni}{II}] lines. The red line indicates the mean flux of all fit models at each wavelength, the orange shaded area marks the $68\,\%$ uncertainty of the fit. }
    \label{OpticalSpectra}
\end{figure*}

Nebular phase spectra of SNe Ia exhibit a number of broad ($\approx7\,000$ to $9\,000$ km\,s$^{-1}$) emission features (see Fig.~\ref{OpticalSpectra}). In the NIR, we identify the strongest features as transitions of singly ionized [\ion{Fe}{II}] and [\ion{Co}{II}]. The 10\,190\,\AA\,(a$^3$F$_4$--b$^3$F$_4$) transition of [\ion{Co}{II}] decreases in strength according to the decay of $^{56}$Co to $^{56}$Fe \citep{2004A&A...426..547S, 2018A&A...620A.200F}. The emission feature at around 13\,000\,\AA\,is identified as the 12\,570\,\AA\,a$^6$D--a$^4$D multiplet of [\ion{Fe}{II}]. The double peaked feature around 16\,000\,\AA\,is composed of a blend of [\ion{Fe}{II}] and [\ion{Co}{II}] lines of the multiplets a$^4$F--a$^4$D and a$^5$F--b$^3$F, respectively. Redwards of the strong telluric absorption feature at $\sim$ 18\,500\,\AA\,we also detect the a$^{2}$F$_{7/2}$--a$^{4}$F$_{9/2}$ line of [\ion{Ni}{II}] in spectra with high SNR \citep{2018A&A...619A.102D}.

In the optical we see blends of singly and doubly ionized Fe, Co and Ni. The strong feature at 4\,700\,\AA\,originates mainly from the $3d^6$\,$^5$D--$^3$F multiplet of [\ion{Fe}{III}] \citep{1980PhDT.........1A, 1994ApJ...426L..89K}. The broad emission centered around $5900$\,\AA\,is primarily due to [\ion{Co}{III}] in the a$^4$F--a$^2$G multiplet. The identification of the $5\,900$\,\AA\,[\ion{Co}{III}] feature is secured by the fact that the relative strength of this feature with respect to e.g. the [\ion{Fe}{III}] $4\,700$\,\AA\,feature decreases with time as predicted by radioactive decay of $^{56}$Co \citep{1994ApJ...426L..89K, 2015MNRAS.454.3816C,2014MNRAS.439.3114D}. Near the $7\,200$\,\AA\,region the spectra exhibit emission lines of the [\ion{Fe}{II}] multiplets a$^4$F--a$^2$G and a$^6$D--a$^4$P and the [\ion{Ni}{II}] multiplet z$^2$D--a$^2$F. The identification of the various emission lines in the optical and NIR of SNe Ia in the nebular phase has been extensively discussed in the literature and is considered secure. A detailed overview of the strongest emission lines is given in Table~\ref{tab:LineID}.
\begin{table}
	\centering
	\caption{Selected forbidden lines of singly and doubly ionized Fe, Co and Ni in the optical and NIR} 
	\label{tab:LineID}
    \begin{tabular}{lcc}
    	\hline 
    	$\lambda_{\text{rest}}$(\AA) & Ion & Transition\\
		\hline  
		4\,418 & [\ion{Fe}{II}] & a$^6$D$_{9/2}\,-\,$b$^4$F$_{9/2}$\\ 
		4\,659 & [\ion{Fe}{III}] & $^5$D$_{4}\,-\,^3$F$_{4}$\\
		4\,891 & [\ion{Fe}{II}] & a$^6$D$_{7/2}\,-\,$b$^4$P$_{5/2}$\\ 
		5\,160 & [\ion{Fe}{II}] & a$^4$F$_{9/2}\,-\,$a$^4$H$_{13/2}$\\ 
		5\,272 & [\ion{Fe}{III}] & $^5$D$_{3}\,-\,^3$P$_{2}$\\
        5\,528 & [\ion{Fe}{II}] & a$^4$F$_{7/2}\,-\,$a$^2$D$_{5/2}$\\
        5\,888 & [\ion{Co}{III}] & a$^4$F$_{9/2}\,-\,$a$^2$G$_{9/2}$\\
        5\,908 & [\ion{Co}{III}] & a$^4$F$_{7/2}\,-\,$a$^2$G$_{7/2}$\\
        6\,197 & [\ion{Co}{III}] & a$^4$F$_{7/2}\,-\,$a$^2$G$_{9/2}$\\
        6\,578 & [\ion{Co}{III}] & a$^4$F$_{9/2}\,-\,$a$^4$P$_{5/2}$\\
        6\,855 & [\ion{Co}{III}] & a$^4$F$_{7/2}\,-\,$a$^4$P$_{3/2}$\\
        7\,155 & [\ion{Fe}{II}] & a$^4$F$_{9/2}\,-\,$a$^2$G$_{9/2}$\\
        7\,172 & [\ion{Fe}{II}] & a$^4$F$_{7/2}\,-\,$a$^2$G$_{7/2}$\\
        7\,378 & [\ion{Ni}{II}] & z$^2$D$_{5/2}\,-\,$a$^2$F$_{7/2}$\\
		7\,388 & [\ion{Fe}{II}] & a$^4$F$_{5/2}\,-\,$a$^2$G$_{7/2}$\\
        7\,414 & [\ion{Ni}{II}] & z$^2$D$_{3/2}\,-\,$a$^2$F$_{5/2}$\\
        7\,453 & [\ion{Fe}{II}] & a$^4$F$_{7/2}\,-\,$a$^2$G$_{9/2}$\\
        7\,638 & [\ion{Fe}{II}] & a$^6$D$_{7/2}\,-\,$a$^4$P$_{5/2}$\\
        7\,687 & [\ion{Fe}{II}] & a$^6$D$_{5/2}\,-\,$a$^4$P$_{3/2}$\\
        8\,617 & [\ion{Fe}{II}] & a$^4$F$_{9/2}\,-\,$a$^4$P$_{5/2}$\\
        9\,345 & [\ion{Co}{II}] & a$^3$F$_{3}\,-\,$a$^1$D$_{2}$\\
        9\,704 & [\ion{Fe}{III}] & $^3$H$_{6}\,-\,^1$I$_{6}$\\
        10\,190 & [\ion{Co}{II}] & a$^3$F$_{4}\,-\,$b$^3$F$_{4}$\\
        10\,248 & [\ion{Co}{II}] & a$^3$F$_{3}\,-\,$b$^3$F$_{3}$\\
        10\,611 & [\ion{Fe}{III}] & $^3$F$_{4}\,-\,^1$G$_{4}$\\
        12\,570 & [\ion{Fe}{II}] & a$^6$D$_{9/2}\,-\,$a$^4$D$_{7/2}$\\
        12\,943 & [\ion{Fe}{II}] & a$^6$D$_{5/2}\,-\,$a$^4$D$_{5/2}$\\
        13\,206 & [\ion{Fe}{II}] & a$^6$D$_{7/2}\,-\,$a$^4$D$_{7/2}$\\
        15\,335 & [\ion{Fe}{II}] & a$^4$F$_{9/2}\,-\,$a$^4$D$_{5/2}$\\
        15\,474 & [\ion{Co}{II}] & a$^5$F$_{5}\,-\,$b$^3$F$_{4}$\\
        15\,488 & [\ion{Co}{III}] & a$^2$G$_{9/2}\,-\,$a$^2$H$_{9/2}$\\
        15\,995 & [\ion{Fe}{II}] & a$^4$F$_{7/2}\,-\,$a$^4$D$_{3/2}$\\
        16\,440 & [\ion{Fe}{II}] & a$^4$F$_{9/2}\,-\,$a$^4$D$_{7/2}$\\
        17\,416 & [\ion{Co}{III}] & a$^2$G$_{9/2}\,-\,$a$^2$H$_{11/2}$\\ 
        17\,455 & [\ion{Fe}{II}] & a$^4$F$_{3/2}\,-\,$a$^4$D$_{1/2}$\\ 
        18\,098 & [\ion{Fe}{II}] & a$^4$F$_{7/2}\,-\,$a$^4$D$_{7/2}$\\ 
        19\,390 & [\ion{Ni}{II}] & a$^2$F$_{7/2}\,-\,$a$^4$F$_{9/2}$\\
        20\,028 & [\ion{Co}{III}] & a$^4$P$_{5/2}\,-\,$a$^2$P$_{3/2}$\\ 
        20\,157 & [\ion{Fe}{II}] & a$^2$G$_{9/2}\,-\,$a$^2$H$_{9/2}$\\ 
        22\,184 & [\ion{Fe}{III}] & $^3$H$_{6}\,-\,^3$G$_{5}$\\
		\hline
	\end{tabular}
\end{table}

\section{Methods}
\label{SectionModels}

\subsection{Summary}
\label{sectionMethodSummary}

We have determined that the line ratio of the 12\,570 to 7\,155\,\AA\,[\ion{Fe}{II}] lines in the nebular spectra of type Ia supernovae evolves with supernova age in a predictable log-linear manner. We assume that evolution is valid for supernovae for which we only have optical coverage. The range of electron densities and temperatures that give rise to a given ratio is determined by the atomic data for these transitions. For each epoch we thus have prior knowledge of the range of n$_e$ and T. This range is used to determine the ratio of the emissivity per atom for the 7\,155 [\ion{Fe}{II}] to 7\,378\,\AA\,[\ion{Ni}{II}] lines and thus determine the range of mass ratios of Nickel to Iron based on optical data alone at any given epoch.

\subsection{The model}
\label{SectionTheModel}

\begin{table}
	\centering
	\caption{Ions included in the fits and their atomic data sets.}
	\label{tab:AtomicData}
    \begin{tabular}{llll}
    	\hline 
    	Ion 			& Levels$^a$ & Ref. $A_{ij}\,^b$   & Ref. $\Upsilon_{ij}\,^c$ \\	
		\hline  
		\text{Fe\,\textsc{\lowercase{II}}}  & 52  & \citet{2015ApJ...808..174B} & \citet{2015ApJ...808..174B} \\
		\text{Fe\,\textsc{\lowercase{III}}} & 39  & \citet{1996AAS..116..573Q} & \citet{1996AAS..119..523Z} \\
		\text{Co\,\textsc{\lowercase{II}}}  & 15  & \citet{2016MNRAS.456.1974S} & \citet{2016MNRAS.456.1974S} \\
		\text{Co\,\textsc{\lowercase{III}}} & 15  & \citet{2016MNRAS.459.2558S} & \citet{2016MNRAS.459.2558S} \\
		\text{Ni\,\textsc{\lowercase{II}}}  & 18  & \citet{2016AA...587A.107C} & \citet{2010AA...513A..55C}  \\
		\text{Ni\,\textsc{\lowercase{III}}} & 9   & \citet{2016AA...585A.121F} & \citet{1998JPhB...31..145W}  \\
		\hline
		\multicolumn{4}{l}{\footnotesize$^a$Energy levels and statistical weights are taken}\\
		\multicolumn{4}{l}{\footnotesize$\,\,\,\,$from NIST \citep{NIST_ASD}.}\\
        \multicolumn{4}{l}{\footnotesize$^b$Einstein $A$ coefficient between levels $i$ and $j$.}\\
        \multicolumn{4}{l}{\footnotesize$^c$Maxwellian averaged collisional strength between levels $i$ and $j$.}\\
	\end{tabular}
\end{table}
We use a one-zone model as described in \citet{2018A&A...620A.200F}. We extend the model to include all first and second ionisation stages of iron, nickel and cobalt (see Table~\ref{tab:AtomicData}). For this set of ions we solve the NLTE rate equations and compute level populations. Throughout this work we redshift and extinguish the spectral models instead of correcting the observed spectra. In Table~\ref{TableSpectraOverview} we show the redshift and reddening applied to our models. For the reddening correction in our models we adopt \citet{1989ApJ...345..245C}. The strength of the reddening is strongly constrained by the presence of a number of lines arising from the same upper level in different ions (e.g. 12\,570 and 16\,440\,\AA\,of [\ion{Fe}{II}]).

We assume that thermal emission is the dominant source of light from the start of the nebular phase until $\sim$ 500 days after the explosion. During this phase, the ejecta are transparent for optical and NIR photons, allowing us to ignore radiative transfer effects. We also do not consider non-thermal excitations as the energy going into this channel at the relatively high electron densities we determine is also very low \citep{1989ApJ...343..323F}. We  do  not  include  charge  exchange and time-dependent terms in the NLTE rate equations. For the set of ions given in Table~\ref{tab:AtomicData} we solve the NLTE rate equations to obtain the level populations of the ions, which are used to determine the line emissivities.

We compare our parameterised model $M$ to the XShooter observations $D$ described in Section~\ref{SectionObservations} using the approach from \citet{2015ApJ...812..128C}. The likelihood function contains a correlation matrix $C$ which has the uncertainties of the pixels as diagonal elements and the correlations of nearby pixels on the off-diagonals:
\small
\begin{equation}
    \ln p(D\vert M) = -\frac{1}{2}\left( (D-M)^\text{T}C^{-1}(D-M) + \ln \det C + N_{\text{pix}}\ln{2\pi}\right)
\end{equation}
\normalsize
To account for systematic imperfections of the model (e.g. line profiles are not Gaussian far from the line center), we use Gaussian processes with a Mat\'ern kernel to add an additional noise term in the correlation matrix at the location of the feature edges \citep[see ][]{2015ApJ...812..128C}. This prevents the sampling algorithm from only choosing a narrow set of parameter values, which yield a better fit in regions where the model is systematically unable to fit the observations. We employ flat priors for all parameters of the model. The upper and lower bounds of the flat priors are chosen in such a way that the posterior parameter distributions are not truncated. 

\begin{figure}
	\includegraphics[width=\linewidth]{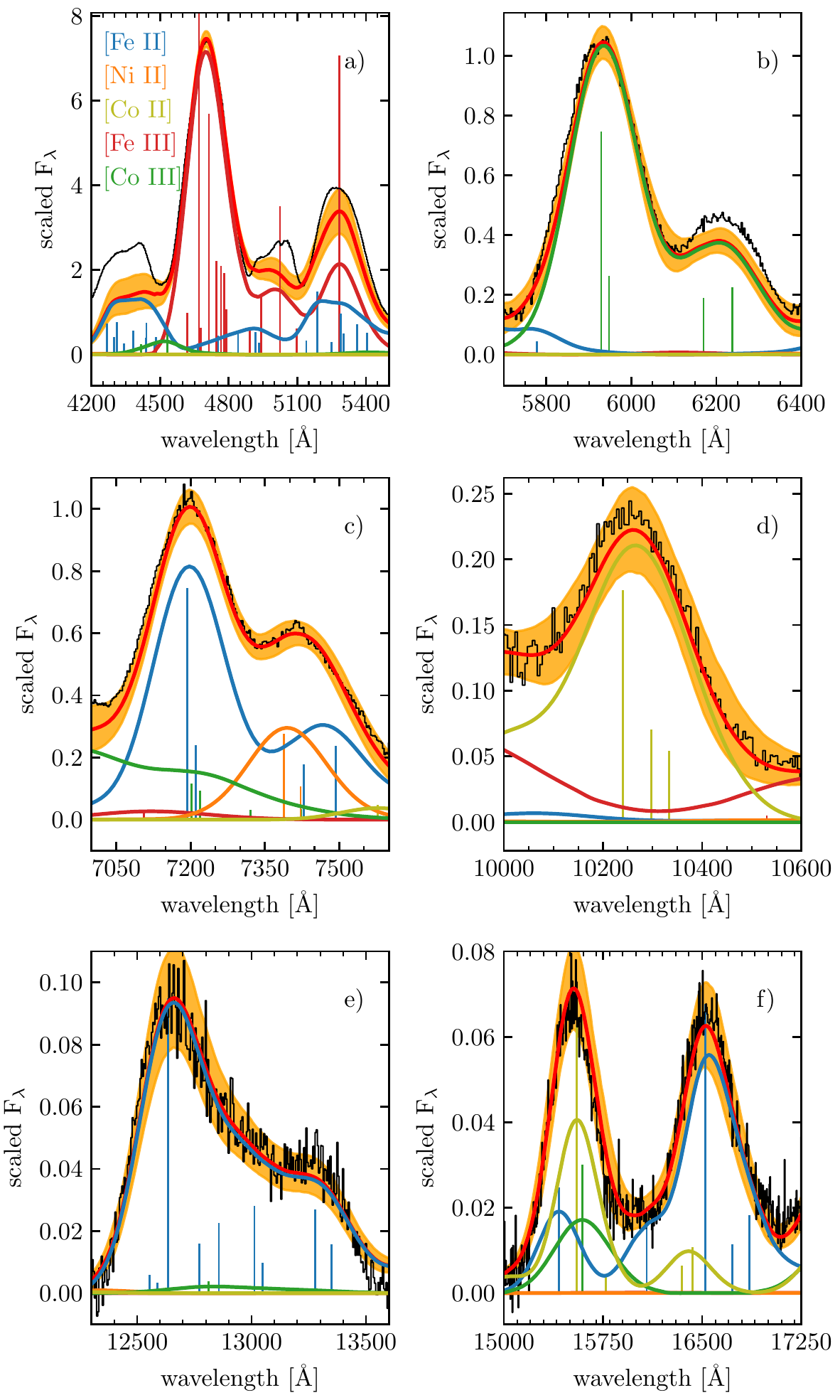}
    \caption{Example fit of SN\,2017bzc at +215\,days after B-band maximum. The individual panels highlight the ionic emission contributions to six features from the near-UV to the NIR. The strongest lines of each ion are indicated as vertical lines. The height of these lines shows their relative strengths. The flux of the spectrum was scaled to the 7\,155\,\AA\,peak. The observed spectrum is not corrected for extinction and redshift of the host. Instead, all model lines are extinguished and redshifted.}
    \label{ExampleFit}
\end{figure}

We use nested sampling to find the posterior distributions of the parameters of the model that yield good fits with the observed spectrum \citep[https://github.com/kbarbary/nestle, see also][]{2007MNRAS.378.1365S}. Fig.~\ref{OpticalSpectra} presents the fits results for the spectra given in Table~\ref{tab:Observations}. The red line indicates the mean flux of all fit models at each wavelength while the orange shaded area marks the 68\% uncertainty of the fit. Fit results for the previously published spectra of the XShooter sample are shown in \citet{2018A&A...620A.200F}. An exemplary zoom into the fit of SN\,2017bzc at $+215\,$days is shown in Fig.~\ref{ExampleFit}.

For each spectrum we can use the posterior distribution of the model parameters to compute line emissivities of all lines of singly and doubly ionized Fe, Ni and Co. In this work we use line ratios of [\ion{Ni}{II}] and [\ion{Fe}{II}]. \ion{Ni}{II} emission in the nebular phase can only be the result of the stable isotope $^{58}$Ni, as the radioactive material has long since decayed. Fe can be produced directly during the explosion as $^{54,56}$Fe or it can be the decay product of radioactive $^{55}$Co, $^{56}$Ni and $^{57}$Ni. The line ratio of [\ion{Ni}{II}] and [\ion{Fe}{II}] allows us to determine the mass fraction of neutron rich (leading to $^{58}$Ni) to radioactive material, which in turn can be compared to predictions of explosion models. A similar study was performed for the NIR line ratio of the 15\,470\,\AA\,[\ion{Co}{II}] to the 15\,330\,\,AA\ [\ion{Fe}{II}] line in \citet{2018A&A...620A.200F}. While the NIR nebular spectra are easier to model than the optical spectra, the mass ratio of \ion{Co}{II} to \ion{Fe}{II} changes with time and the number of spectra with NIR coverage is quite limited. In this work we want to make use of several decades of optical nebular phase spectroscopy to determine the distribution of the Ni/Fe abundance and compare our findings with predictions from explosion models.

\subsection{Calibration of optical spectra of SNe Ia}
\label{Calibration}
\begin{figure}
	\includegraphics[width=\linewidth]{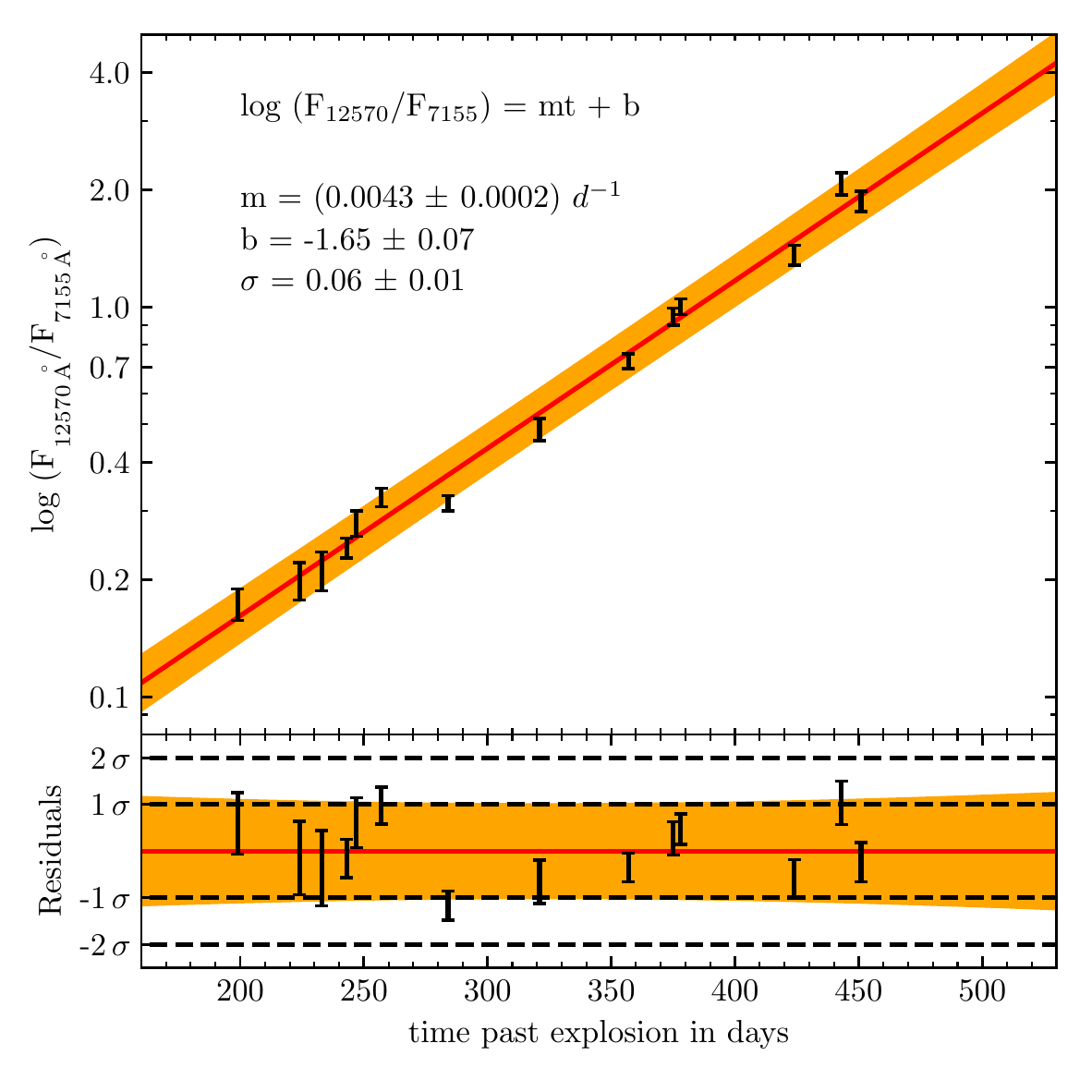}
    \caption{Inferred ratio of the \text{Fe\,\textsc{\lowercase{II}}} 12\,570\,\AA\,to 7\,155\,\AA\,lines as a function of time after explosion. The red line marks a linear fit to data of the form y = mt + b with intrinsic scatter $\sigma$. The orange shaded band indicates the $68\%$ confidence interval of the regression curve.}
    \label{FigureRatioNIR_VIS}
\end{figure}

To determine the \ion{Ni}{II}\,/\,\ion{Fe}{II} mass ratio we compute the ratio of the $7\,378$\,\AA\,[\ion{Ni}{II}] and the 7\,155\,\AA\,[\ion{Fe}{II}] lines (see Fig.~\ref{ExampleFit} panel c). The conversion of line emissivities to emitting masses requires knowledge of the temperature and density of the emitting material. The one-zone-model employed in this study does not allow us to disentangle these two parameters. However, we find that the evolution of the ratio of the strongest \ion{Fe}{II} line in the NIR (12\,570\,\AA) and optical (7\,155\,\AA) is very similar across our sample of optical+NIR spectra (see Fig.~\ref{ExampleFit} panel c and e for these lines). This seems to be a natural evolution from high temperatures and high densities towards lower values. Due to the decreasing temperature it becomes more difficult at late epochs to excite the levels giving rise to optical transitions, thus increasing the ratio of the NIR to optical lines. We fit a simple linear relation through our inferred data points (see Fig.~\ref{FigureRatioNIR_VIS}). The uncertainties of the individual data points are uncorrelated, thus justifying the use of a simple Chi-Square likelihood
\small
\begin{equation}
    \ln p(y|t, \Delta y, m, b, \sigma) = -\frac{1}{2}\sum_n \left( \frac{(y_n -mx_n -b)^2}{s_n^2} + \ln(2\pi s_n^2)\right)
\end{equation}
\normalsize
where
\begin{equation}
    s_n^2 = \sigma_n^2 + \sigma^2(mx_n+b)^2.
\end{equation}
In this equation y and $\Delta y$ indicate the inferred values and uncertainties of the \ion{Fe}{II} 12\,570\AA\,to 7\,155\,\AA\,ratio for our sample, $m$ is the slope of the fit curve, $b$ is its intersect, and $\sigma$ is the intrinsic scatter of the population.
We add an intrinsic scatter term to the likelihood function that takes into consideration that our sample consists of many different objects. The uncertainty of the fit is then a combination of the uncertainty of slope and intersect and the intrinsic scatter term.
We find for the ratio of \ion{Fe}{II} 12\,570\AA\,to 7\,155\,\AA
\small
\begin{equation}
    \label{EquationNIROPT}
    \log\frac{F_{12\,570}}{F_{7\,155}} = -(1.65\pm0.07) + (0.0043\pm0.0002)\,d^{-1}\,\times t_{\text{exp}}[\text{days}]
\end{equation}
\normalsize
with an intrinsic scatter of 0.06\,dex around the best fit curve. The choice of the atomic data has only very weak consequences on the inferred NIR/VIS ratio. Translating the NIR/VIS ratio to temperatures/densities does rely on the atomic data, however. The atomic data used throughout this work is given in Table~\ref{tab:AtomicData}.

\begin{figure}
	\includegraphics[width=\linewidth]{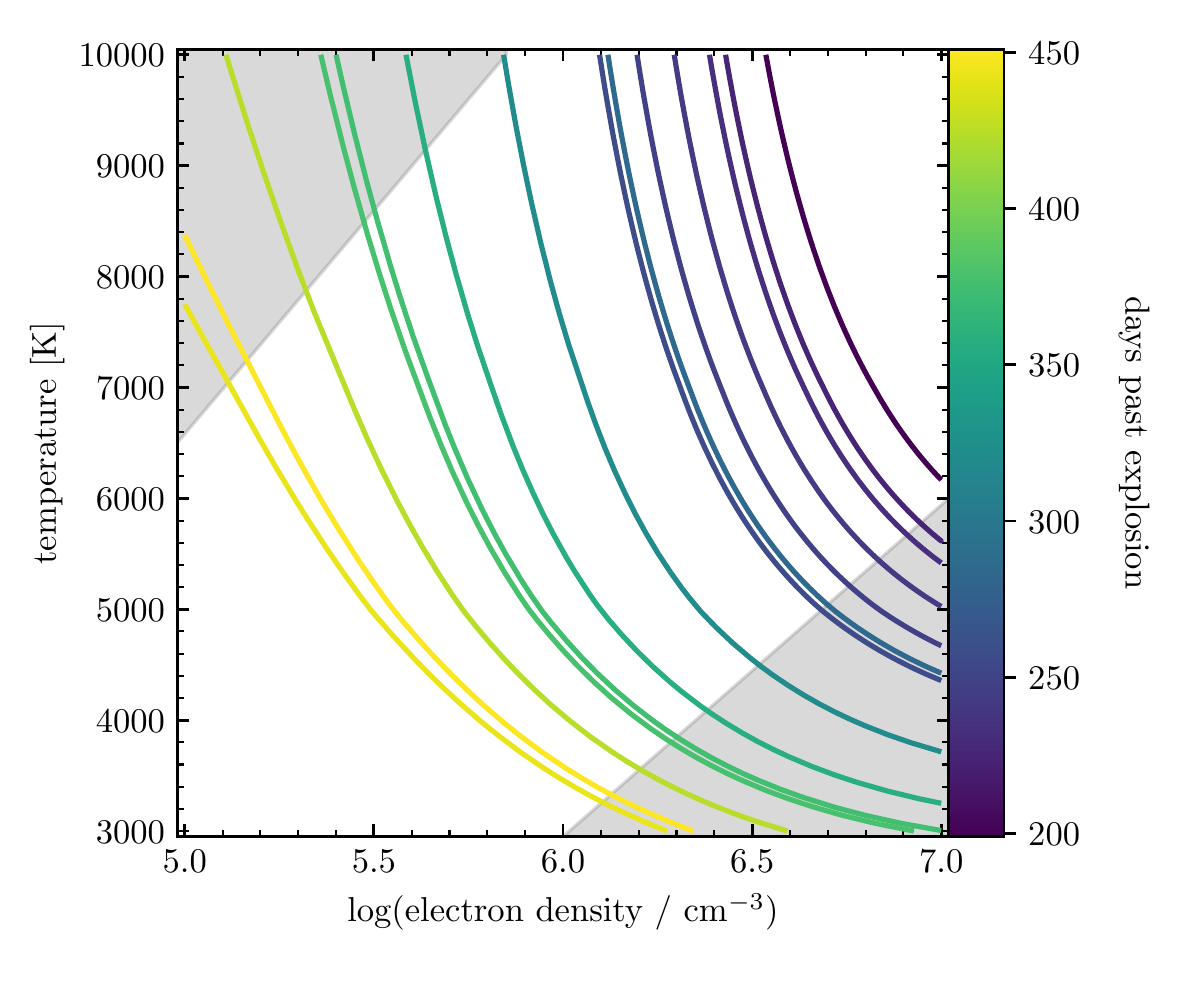}
    \caption{Allowed regions of the electron density and temperature for the SN Ia in our sample. Every curve corresponds to one value of the 12\,570\,\AA\,to 7\,155\,\AA\,\ion{Fe}{II} line ratio. The allowed region is evolving with time to lower temperatures and densities. The density evolves as $t^{-3}$ in accordance with homologous  expansion of the ejecta. Colors indicate the epochs of the spectra. The grey shaded regions (high density + low temperature; low density + high temperature) are excluded by the fits to [\ion{Ni}{II}] and [\ion{Co}{II}].}
    \label{FigureTemperatureDensity}
\end{figure}

Alone, the optical spectra of SNe Ia do not allow us to constrain the temperature and density of the emitting material in any meaningful way - we can obtain good fits for a wide range of temperatures and densities. However, we notice that for a given \ion{Fe}{II} 12\,570\AA\,to 7155\,\AA\,ratio only specific tracks in the temperature/density space are possible. The inference uncertainty of the NIR/VIS ratio translates into a curve with non-zero width in the temperature/density space. The measurement of the NIR/VIS line ratio is considered robust - no other strong lines are present in the 12\,500\,\AA\,feature, and in the 7\,000\,\AA\,region only \ion{Co}{III} of the iron group elements has a weak contribution. We exclude the extremes in the temperature/density space (see grey shaded areas in Fig.~\ref{FigureTemperatureDensity}) by fitting the many lines of singly and doubly ionized material at optical and NIR wavelengths. Each of the curves in Fig.~\ref{FigureTemperatureDensity} corresponds to one value of the NIR/VIS ratio. We can thus determine a range of temperatures and densities of SNe Ia in the nebular phase assuming that the \ion{Fe}{II} 12\,570\AA\,to 7\,155\,\AA\,ratio evolves as the red curve in Fig.~\ref{FigureRatioNIR_VIS} with a 1-sigma uncertainty of 0.06\,dex. We add this constraint as a Gaussian prior into the likelihood function of our Bayesian fit model. 

\subsection{Determination of the Ni to Fe ratio}
The temperature range is significant and thus the uncertainty in the absolute masses is large. A temperature difference of only a few hundred Kelvin can lead to an emitting mass that is different by a factor of a few. However, a more robust quantity is the mass ratio of ions of the same ionisation stage. Under the assumption that the same emitting region gives rise to these lines  \citep{2018MNRAS.477.3567M,2018A&A...620A.200F} the physical conditions of the ions (temperature and electron density) are similar. Using a mass ratio also negates the effect of the rather unknown distance to the SN host galaxy and significantly reduces the effect of the emitting temperature. 

For a given temperature and density we can directly infer the ratio of the number of emitting \ion{Fe}{II} and \ion{Ni}{II} ions required to match the observed flux ratio of the 7\,155\,\AA\,and 7\,378\,\AA\,lines (see Fig.~\ref{ExampleFit} panel c). Temperatures and densities that yield a good fit can be found if a NIR spectrum is available. For spectra that lack this additional information we have to use the relation obtained in Section~\ref{Calibration}. We discuss the additional uncertainties from using the fit relation instead of the full optical + NIR spectrum in Section~\ref{NiFeTest}.

\begin{table}
	\centering
	\scriptsize
	\caption{Results of the ratio of the 12\,570\,\AA\,and 7\,155\,\AA\,lines of \ion{Fe}{II}, the M$_{\text{Co}}$\,/\,M$_{\text{Fe}}$ ratio and the M$_{\text{Ni}}$\,/\,M$_{\text{Fe}}$ ratio for the extended XShooter sample.}
	\label{tab:LineRatio}
    \begin{tabular}{lclccc}
    	\hline 
    	SN 		  & Ref$^a$ & Epoch & R$_{12570/7155}$ &	M$_{\text{Co}}$/M$_{\text{Fe}}$ & M$_{\text{Ni}}$/M$_{\text{Fe}}$\\
		\hline  
		SN\,2015F   & TW & +181\,d & $0.173\pm0.16$ & 0.231$\pm$0.02 & 0.061$\pm$0.010\\
    	PSNJ1149  & M18 & +206\,d & $0.199\pm0.022$ & 0.152$\pm$0.012$^b$ & 0.044$\pm$0.011\\
    	SN\,2017bzc & TW & +215\,d & $0.211\pm0.044$ & 0.154$\pm$0.01 &0.035$\pm$0.009\\
        SN\,2015F   & TW & +225\,d & $0.241\pm0.018$ & 0.142$\pm$0.02 & 0.055$\pm$0.008\\
        SN\,2013ct  & M16 & +229\,d & $0.279\pm0.021$ & 0.103$\pm$0.010$^b$ & 0.037$\pm$0.006\\
        SN\,2015F   & TW & +239\,d & $0.325\pm0.022$ & 0.127$\pm$0.02 & 0.052$\pm$0.008\\
        SN\,2015F   & TW & +266\,d & $0.314\pm0.020$ & 0.097$\pm$0.01 & 0.055$\pm$0.009\\
        SN\,2013cs  & M16 & +303\,d & $0.486\pm0.029$ & 0.066$\pm$0.011$^b$ & 0.031$\pm$0.006\\
        SN\,2012cg  & M16 & +339\,d & $0.726\pm0.021$ & 0.051$\pm$0.005$^b$ & 0.038$\pm$0.006\\
        SN\,2012fr  & M16 & +357\,d & $0.947\pm0.048$ & 0.038$\pm$0.004$^b$ & 0.025$\pm$0.005\\
    	SN\,2013aa  & M16 & +360\,d & $1.00\pm0.068$ & 0.035$\pm$0.003$^b$ & 0.033$\pm$0.006\\
        SN\,2015F   & TW & +406\,d & $1.36\pm0.06$ & $-$ & 0.049$\pm$0.009\\
        SN\,2013aa  & M18 & +425\,d & $2.07\pm0.14$ & 0.025$\pm$0.003$^b$ & 0.035$\pm$0.007\\
        SN\,2012ht  & M16 & +433\,d & $1.87\pm0.11$ & 0.020$\pm$0.005 & 0.009$\pm$0.004\\
		\hline
	\end{tabular}
	\begin{flushleft}
         $^a$Source of the nebular spectrum: TW (This Work); M16 \citep{2016MNRAS.457.3254M}; M18 \citep{2018MNRAS.477.3567M}
         $^b$Result taken from \citet{2018A&A...620A.200F}
    \end{flushleft}
\end{table}
\begin{figure}
	\includegraphics[width=\linewidth]{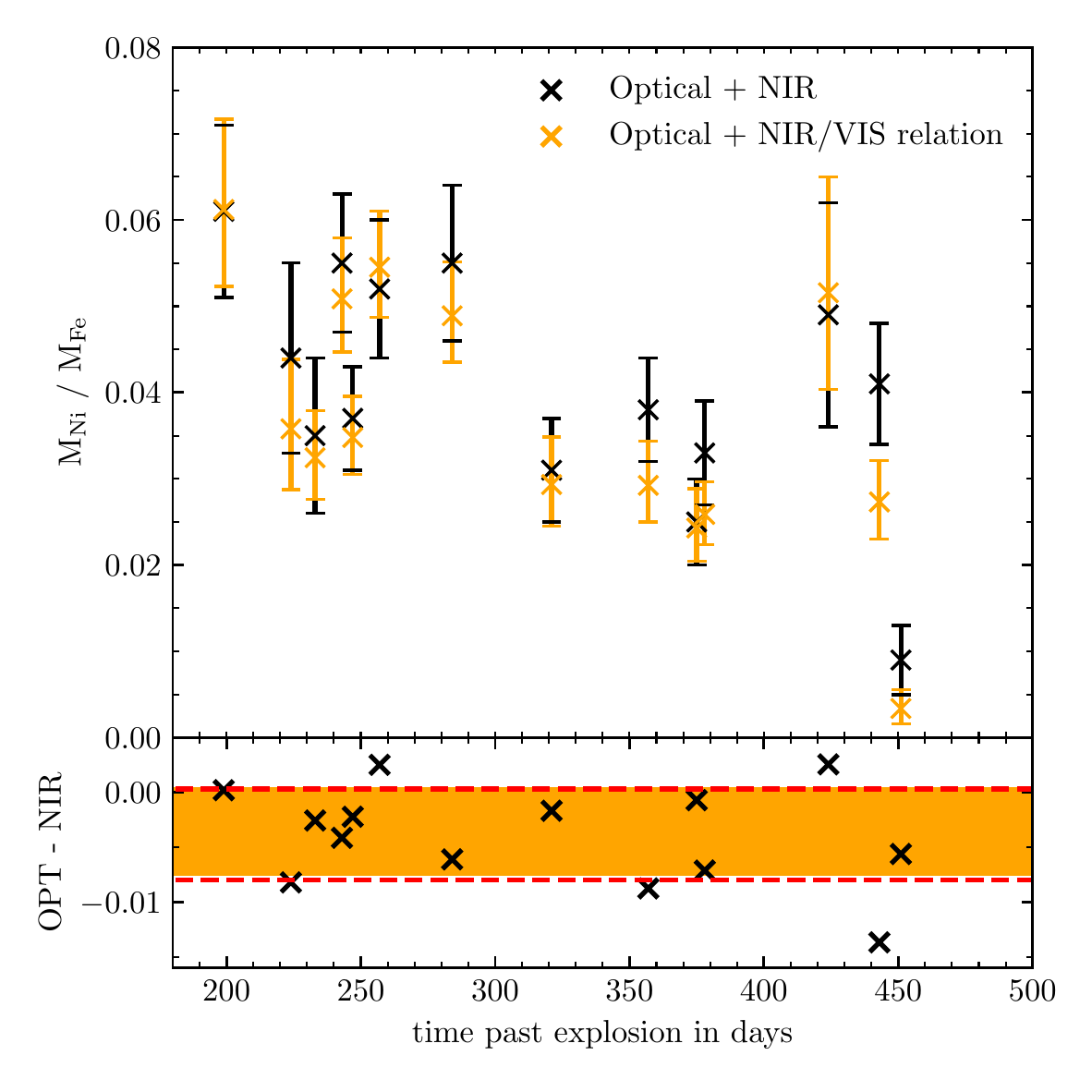}
    \caption{Inferred mass ratio of \text{Ni\,\textsc{\lowercase{II}}} and \text{Fe\,\textsc{\lowercase{II}}} from optical and NIR spectroscopy. Black data points indicate that the \text{Fe\,\textsc{\lowercase{II}}} NIR/VIS ratio was directly inferred from a spectrum covering 4,000--20,000\,\AA. Orange data points indicate that only the optical part of the spectrum was used in conjunction with the relation from Fig.~\ref{FigureRatioNIR_VIS} as a prior. We assume a rise time of $\sim 18$ days \citep{2011MNRAS.416.2607G} to compute the time after explosion. The bottom panel shows the systematic differences between the two methods -- optical spectra + the NIR/VIS relation (OPT) and fitting the full spectrum (NIR). The orange shaded band in the bottom panel marks the 68\% confidence interval of the systematic uncertainty $\sigma_{\text{sys}}=-0.0033^{+0.0037}_{-0.0041}$.}
    \label{NiOverFe}
\end{figure}
\begin{figure*}
	\includegraphics[width=\linewidth]{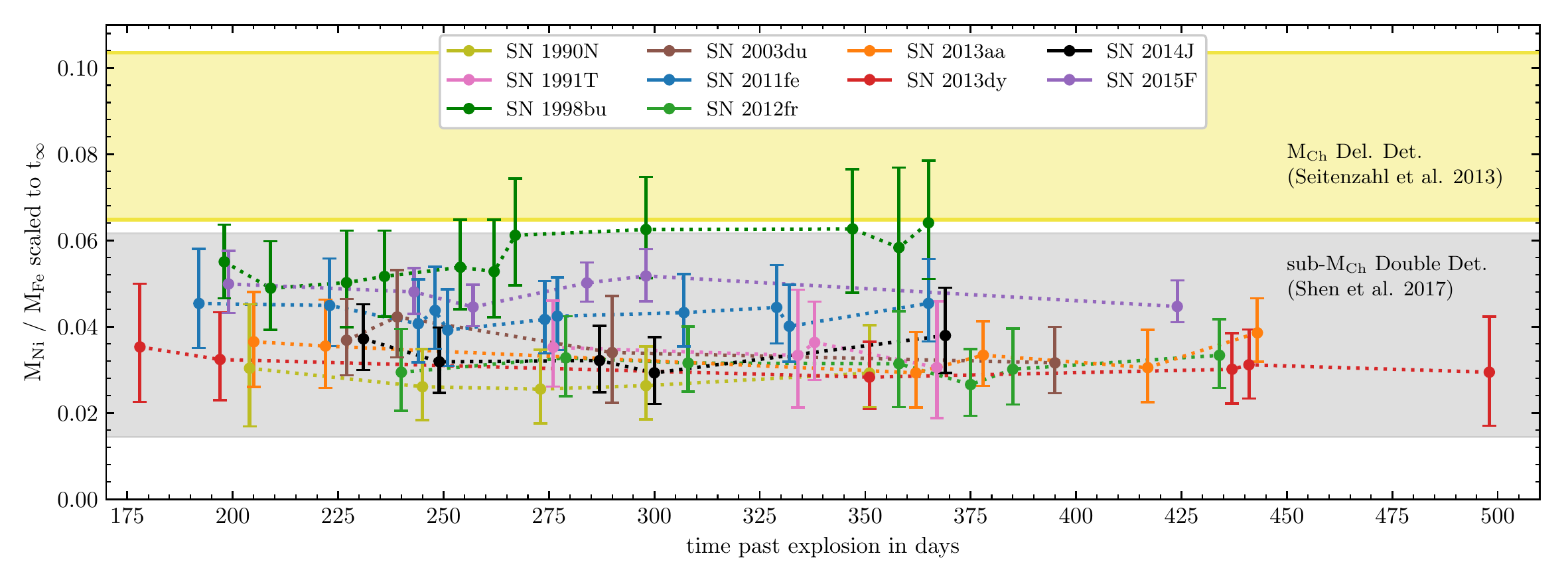}
    \caption{Inferred mass ratio of Ni and Fe for supernovae with multiple observations between $\sim$\,200 and 500 days after the explosion. Explosion model predictions and inferred data points were scaled to the Ni/Fe abundance at $t\rightarrow\infty$ to remove the time dependence of M$_{\text{Fe}}(t)$ in order to better illustrate the consistency of the method. We assume a rise time of $\sim 18$ days \citep{2011MNRAS.416.2607G} to compute the time after explosion. Same colors indicate multiple observations of a supernova. Error bars represent the combined statistical fitting uncertainty and the systematic uncertainty from using the NIR/VIS relation if the spectrum only covers the optical wavelength region up to 10\,000\,\AA.}
    \label{NiOverFe_test_timesequences}
\end{figure*}

\section{Discussion}
\subsection{The Fe$\,\textsc{II}$ NIR/VIS ratio}
In Section~\ref{Calibration} we derived a relation between two of the strongest \ion{Fe}{II} lines that are observed in nebular spectra of SNe Ia. Our extended XShooter sample now contains 14 spectra of 9 different SNe. The ratio of the NIR 12\,570\,\AA\,and the 7\,155\,\AA\,lines evolves similarly for all objects in our sample. In physical terms, the ratio of these lines is a direct measure of the cooling and expanding Fe-rich ejecta. The relation does not depend on the collision strengths but only on the transition rates of \ion{Fe}{II}. These are well known, as can be seen from the match of the fit models and the observed spectra in regions where only \ion{Fe}{II} emission is present. Additionally, the \ion{Fe}{II} NIR/VIS relation as presented in this work is not just the result of a possible oversimplification of our one-zone model. It is obtained by effectively de-blending the lines of singly and doubly ionized iron, nickel and cobalt. It only depends on the total emission through the two lines. The assumed Gaussian line profile used in this work only has a marginal effect on the inferred values. More sophisticated explosion multizone models should be able to reproduce the relation by integrating the flux of the 12\,570\,\AA\,and the 7\,155\,\AA\,lines over all emitting regions.

\subsection{Robustness of the Ni/Fe ratio}
\label{NiFeTest}
\subsubsection{Fitting optical XShooter spectra with the Fe$\,\textsc{II}$ NIR/VIS fit relation}
\label{TestNIRVIS_relation}
We test for the presence of systematic effects arising from our method by applying it to the optical component of the Xshooter sample. The results of this comparison study (full spectrum vs optical only) are shown in Fig.~\ref{NiOverFe}. The use of the NIR/VIS relation as a prior does not imply that the posterior of the 12\,570\,\AA\,to 7\,155\,\AA\,line ratio for a given epoch $t_i$ has the same width as the fit curve in Fig.~\ref{FigureRatioNIR_VIS}. In general, fitting the optical spectrum with the use of the NIR/VIS relation does not necessarily prefer the same ratio as fitting the full optical and NIR spectrum. As a result, we obtain different posteriors for the density and temperature for the two fitting methods. It seems that the optical is more sensitive to different regimes of the electron density and temperature than the combined optical and NIR spectrum. On average, the use of the \text{Fe\,\textsc{\lowercase{II}}} 12\,570\,\AA\,to 7\,155\,\AA\,fit relation instead of the NIR spectrum leads to a systematic difference of $\sigma_{\text{sys}}=-0.0033^{+0.0037}_{-0.0041}$. The use of the NIR/VIS relation therefore results in mostly smaller M$_\text{Ni}$\,/\,M$_\text{Fe}$ ratios by about 0.0033 within the 68\% confidence interval. We consider this a systematic uncertainty that adds to the statistical uncertainty linearly.

\subsubsection{Time evolution of the Ni/Fe ratio}
Even though the amount of $^{58}$Ni produced in the explosion is fixed for a single object, the ratio of Ni/Fe changes with time (Fe being the daughter product of $^{56}$Co decay, which at early times has not completely decayed yet). Only after $\sim300$\,days (4 $\times$ t$_{1/2, ^{56}\text{Co}\rightarrow^{56}\text{Fe}}$) the Ni/Fe ratio remains almost constant. 

For supernovae that have several observations during the nebular phase we can test whether our modelling yields consistent Ni to Fe ratios (i.e. that the slope of the data points follows a single theoretical explosion model prediction). In Fig.~\ref{NiOverFe_test_timesequences} we normalize the Ni to Fe to the value at $t=t_\infty$ to make it easier for the reader to see the slope of the measured data points. A flat series of data points indicates that the evolution with time behaves according to the expected yields from the radioactive decay of $^{56}$Ni. For objects with both optical and near-infrared data the full spectrum is fit while for objects with optical data only the method described herein is used to provide the range and evolution of n$_e$ and T.

The evolution of the Ni to Fe mass ratio for objects with multiple observations during the nebular phase is consistent with pure radioactive decay within the statistical uncertainties. A much shallower or steeper slope of the NIR/VIS ratio would lead to non-flat evolutionary curves of the Ni/Fe ratio. 
The only object that shows an evolution of the scaled Ni/Fe mass ratio is SN\,1998bu. Roughly 270 days past its B-band maximum the inferred Ni/Fe mass ratio increases by about 15\% and settles on this new value for the remaining observations. Such a behaviour could be the result of a light-echo contribution to the nebular spectrum, as was found for SN\,1998bu by \citet{2001ApJ...549L.215C}. 

\subsection{A possible contribution of Calcium at 7200\,\AA?}
\label{Catest}
The method presented in Section~\ref{SectionModels} relies on the assumption that only [\ion{Fe}{II}] and [\ion{Ni}{II}] contribute to the 7\,200\,\AA\,feature. If emission from another ion (e.g. \ion{Ca}{II}]) contributes substantially to this feature, our measurement will be systematically wrong as the true contribution of Ni to the feature is lower than estimated from our model. Some NLTE radiative transfer calculations of SNe Ia in the nebular phase predict a non-negligible flux of \ion{Ca}{II}] emission at $\lambda\lambda$ 7\,291.5, 7\,323.9 \citep{2017ApJ...845..176B,2019arXiv190601048W}. If the emitting region is a spherical shell at high velocities outside the iron core, the profile would be flat-topped. Such a plateau of \ion{Ca}{II}] emission would raise the overall flux level in the 7\,200\,\AA\,region without changing the characteristic double peaked shape of the feature.

We can test whether there is a contribution from other ions by fixing the strength of the \text{Ni\,\textsc{\lowercase{II}}} 7\,378\,\AA\,through the 19\,390\,\AA\,line. The relative strength of the two lines only depends on the extinction and the ratio of the transition rates, as they originate from the same upper level:
\begin{equation}
    \frac{F_{19\,390\,\text{\AA}}}{F_{7\,378\,\text{\AA}}} = \frac{A_{^2F_{7/2} - ^4F_{9/2}}(E_{^2F_{7/2}} - E_{^4F_{9/2}})}{A_{^2F_{7/2} - ^2D_{5/2}} (E_{^2F_{7/2}} - E_{^2D_{5/2}})} = 0.202
\end{equation}
The observed strength will depend on the extinction. Unfortunately, there is a strong telluric absorption band just bluewards of the 19\,390\,\AA\,line of [\ion{Ni}{II}]. The SNR in this region is only sufficiently high for a small number of objects in our XShooter sample. \citet{2018A&A...619A.102D} investigated the [\ion{Ni}{II}] 19\,390\,\AA\,line for the nearby SN\,2014J.

The observations of SN\,2015F, one of the closest SNe in the last decade, can be used to further verify this method. We obtained 5 nebular phase XShooter spectra between +181 and +406 days after B-band maximum. The first four epochs (+181, +225, +239, +266 days after maximum) are of exceptional quality and clearly show the 19\,390\,\AA\,line. The observation at +406 days has a SNR that is insufficient to detect such a weak line, especially as it lies close to a telluric feature. SN\,2017bzc was farther away than SN\,2015F, but with an integration time of 10\,080\,s the [\ion{Ni}{II}] 19\,390\,\AA\,line can be seen in the +215\,d spectrum.

\begin{figure*}
	\includegraphics[width=\linewidth]{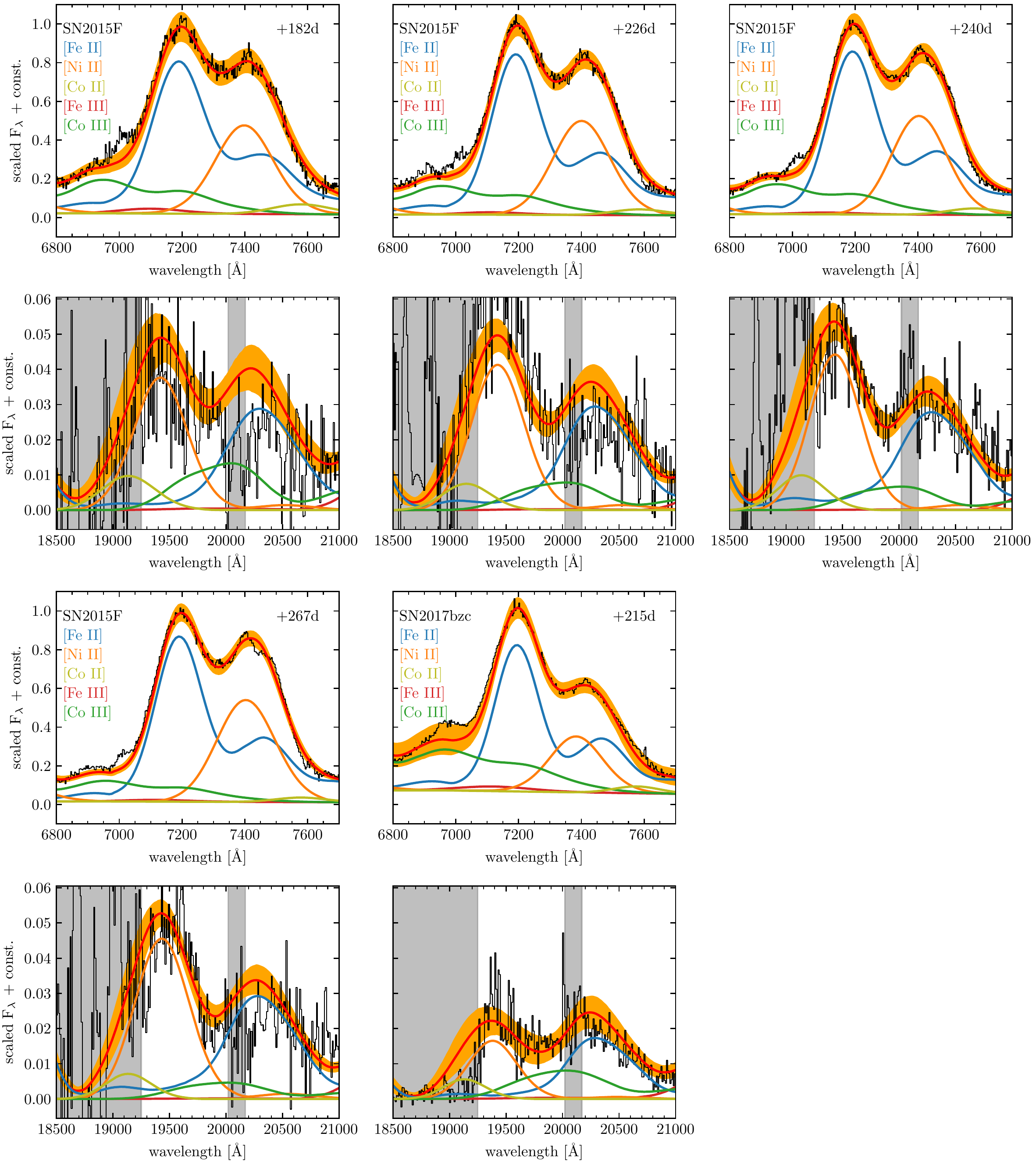}
    \caption{Comparison between the \text{Fe\,\textsc{\lowercase{II}}} and \text{Ni\,\textsc{\lowercase{II}}} dominated regions in the optical at 7\,200\,\AA\,(top panel) and the NIR at $20\,000$\,\AA\,(bottom panel) for four observations of SN\,2015F and one spectrum of SN\,2017bzc. The \text{Ni\,\textsc{\lowercase{II}}} lines at 7\,378\,\AA\,and 19\,390\,\AA\,originate in the same upper level a$^2$F$_{7/2}$ and have therefore a fixed line strength ratio that only depends on the ratio of their transition rates. For the atomic data adopted in this work the ratio of the 19\,390\,\AA\,to the 7\,378\,\AA\,line is 0.202. In the plots the ratio of the two \ion{Ni}{II} emission features is different because of three effects: The optical \ion{Ni}{II} feature is a blend of several lines, the flux density is lower at longer wavelengths and the ratio depends on galactic as well as host galaxy reddening. Regions of extremely low atmospheric transmission are shaded in grey.}
    \label{SN2015F_NiII}
\end{figure*}

An overview of the model fits for each of these spectra is shown in Fig.~\ref{SN2015F_NiII}. The 19\,000\,\AA\,feature has not been used to compute the fits. A significant contribution of \text{Ca\,\textsc{\lowercase{II}}} in the optical would lead to a much weaker 19\,390\AA\,line, which is in contradiction to our observations. A weak \text{Ca\,\textsc{\lowercase{II}}} contribution cannot be ruled out but its effect on the Ni/Fe mass ratio would be very limited. None of the objects with sufficiently high SNR in the 19\,000\,\AA\,region require any \ion{Ca}{II}]. While it is not impossible that some SNe Ia -- transitional objects such as the 86G-like or the faint 91bg-like -- exhibit Calcium emission in the 7\,200\,\AA\,region, the feature can be explained by only [\ion{Fe}{II}] and [\ion{Ni}{II}] for the normal and luminous population of SNe Ia \citep[see also][]{2017MNRAS.472.3437G}.

\subsection{M$_\text{Co/Fe}$ from the extended XShooter sample}
The additional observations can be used to further the work described in \citet{2018A&A...620A.200F}. As has been noted by several authors \citep{2018MNRAS.477.3567M, 2018A&A...620A.200F}, the singly ionized lines of Fe, Ni and Co exhibit the same line shift and width. The same holds true for the two additional SNe with nebular phase XShooter observations presented in this work. It is therefore reasonable to assume that the singly ionized species are co-located in the ejecta and share the physical excitation conditions -- temperature and density. Our updated model allows us to directly compute the Co to Fe mass ratio without having to use LTE approximations. The effect, however, is quite limited for the NIR lines in question ($<\,5\%$). 

\begin{figure*}
	\includegraphics[width=\linewidth]{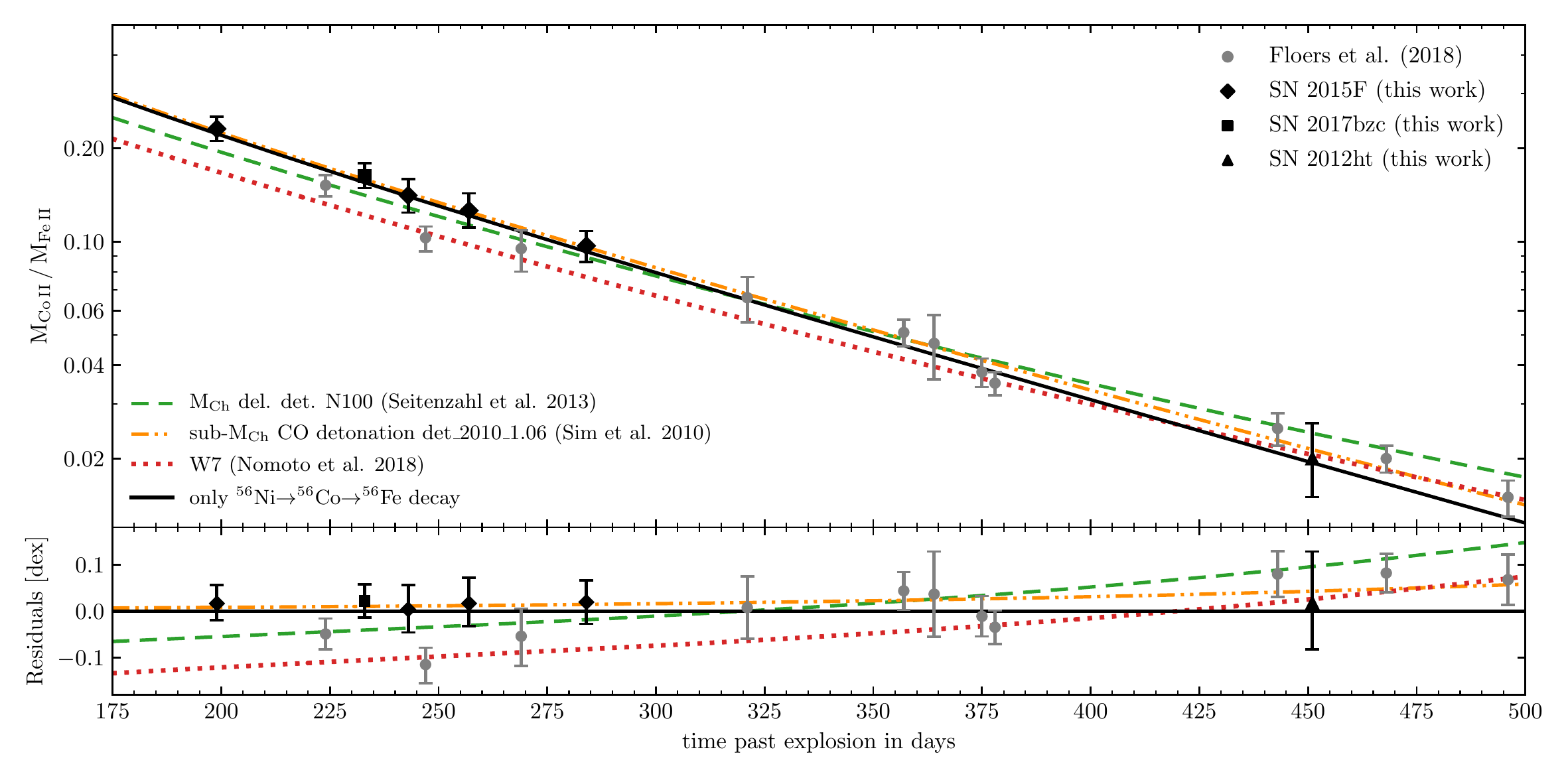}
    \caption{Evolution of the inferred $\mathrm{M_{\ion{Co}{II}}\,/\,M_{\ion{Fe}{II}}}$ ratio with time for the extended XShooter sample. We assumed a rise time of 18 days \citep{2011MNRAS.416.2607G}. The error bars reflect the $68\%$ posterior interval of the mass ratio. The colored lines show the expected mass ratio $\mathrm{M_{Co}\,/\,M_{Fe}}$ of the M$_{\text{Ch}}$ delayed-detonation model `N100' \citep[][green]{2013MNRAS.429.1156S}, the sub-M$_{\text{Ch}}$ CO detonation model `det\_2010\_1.06' \citep[][orange]{2010ApJ...714L..52S} and the M$_{\text{Ch}}$ `W7 Z$_\odot$' model \citep[][red]{2018SSRv..214...67N}. The black line is not a fit to the data and represents the $\mathrm{M_{Co}\,/\,M_{Fe}}$ ratio assuming only radioactive decay from $^{56}$Ni to $^{56}$Co to $^{56}$Fe. Grey data points are from \citet{2018A&A...620A.200F}. Black data points are from the newly published objects in this work (SN\,2015F and SN\,2017bzc). The bottom panel shows the residuals normalized to the pure $^{56}$Ni to $^{56}$Co to $^{56}$Fe decay.}
    \label{CoOverFe}
\end{figure*}

Fig.~\ref{CoOverFe} displays a comparison of the new observations with the ones from \citet{2018MNRAS.477.3567M}. We find that the three new objects (SN\,2012ht, which was not included in the sample of \citeauthor{2018A&A...620A.200F} \citeyear{2018A&A...620A.200F}, SN\,2015F, SN\,2017bzc) have a M$_{\text{Co}}$\,/\,M$_{\text{Fe}}$ ratio that is consistent with sub-M$_{\text{Ch}}$ explosions. Only the spectrum of SN\,2012ht allows us to probe the $^{57}$Ni content in the ejecta as all other spectra are significantly younger than 300 days. For them, the ratio instead is a measure of the fraction of stable iron ($^{54,56}$Fe) to radioactive iron ($^{56}$Ni decay products). 
\subsection{M$_\text{Ni/Fe}$ from archival optical spectra}
\begin{figure*}
	\includegraphics[width=\linewidth]{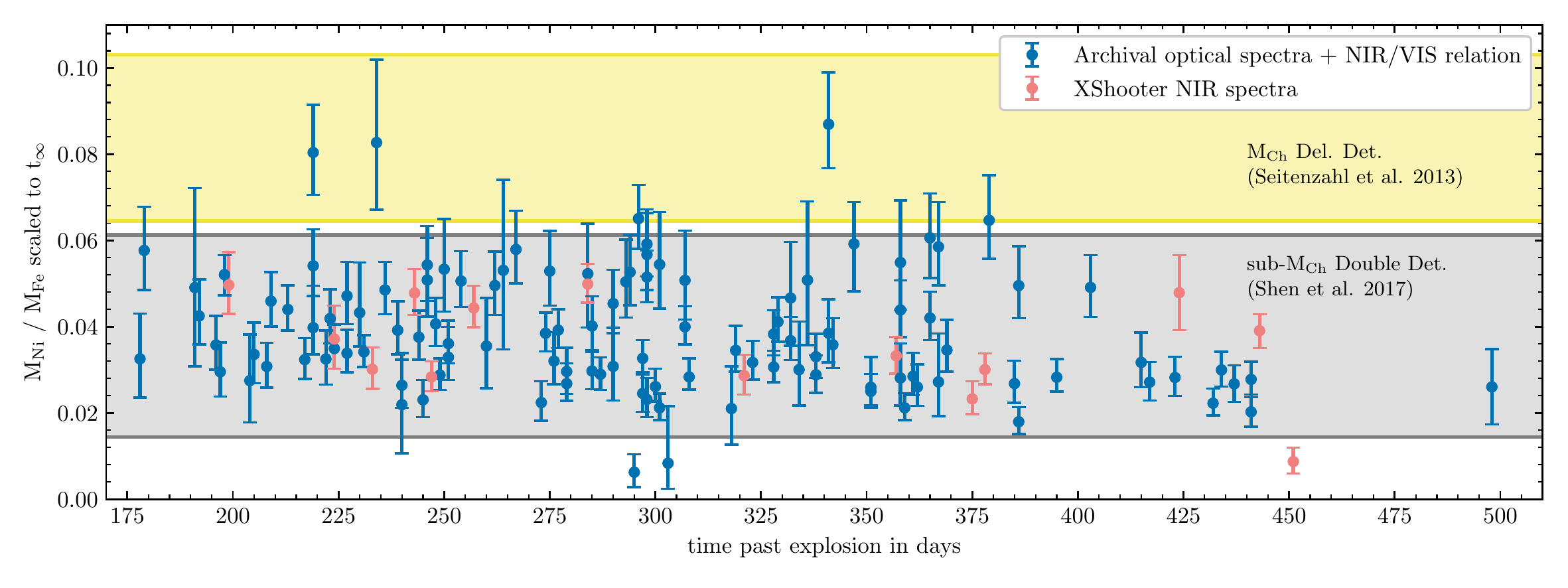}
    \caption{Inferred mass ratio of Ni and Fe from archival optical spectra and XShooter observations. Orange datapoints indicate that the \text{Fe\,\textsc{\lowercase{II}}} NIR/VIS ratio was directly inferred from a spectrum covering 4,000--20,000\,\AA. Blue data points indicate optical nebular phase spectra that have been modelled using the relation from Fig.~\ref{FigureRatioNIR_VIS} as a prior. Errorbars only indicate the statistical uncertainty from the fit. We assume a rise time of $\sim 18$ days \citep{2011MNRAS.416.2607G} to compute the time after explosion. The shaded bands display predictions of the Ni to Fe mass ratio from explosion model simulations \citep{2013MNRAS.429.1156S, 2018ApJ...854...52S}. Inferred and predicted mass ratios were scaled to $t\rightarrow\infty$.}
    \label{NiOverFe_all}
\end{figure*}
The evolution of the NIR/VIS lines of \ion{Fe}{II} allows us to model nebular spectra that cover only the optical wavelength range. We collected 130 spectra of 58 SNe Ia at epochs $>\,170$ days after B-band maximum that have adequate SNR. A full list of all observations used for this study is given in Table~\ref{TableSpectraOverview}. 
The spectra are modelled as described in Section~\ref{SectionModels}. For SNe which have multiple observations in the nebular phase we combine the inferred mass ratios. We report the inferred scaled Ni/Fe mass ratio in Table~\ref{TableSpectraOverview}. No corrections (e.g. fitting optical + NIR spectra vs only optical spectra; Section~\ref{TestNIRVIS_relation}) have been applied to the inferred values. An overview of all objects (XShooter + archival) in our sample is given in Fig.~\ref{NiOverFe_all}. We find that the majority of SNe exhibit Ni/Fe mass ratios below 0.05.

A similar study was conducted by \citet{2018MNRAS.477.3567M} for 8 objects in their XShooter sample. The same objects are also included in this work, however, a different method for the determination of the abundance ratio is used. Instead of modelling the full spectrum, \citet{2018MNRAS.477.3567M} restrict themselves to the $7\,200\,$\AA\,[\ion{Fe}{II}] and [\ion{Ni}{II}] dominated region. To convert the ratio of the LTE line fluxes to an abundance ratio of nickel and iron, they use average departure coefficients of a W7 model \citep{1984ApJ...286..644N,2018SSRv..214...67N} at 330\,days from \citet{2015ApJ...814L...2F}. As this model does not allow a determination of the temperature of the emitting material, \citet{2018MNRAS.477.3567M} assume temperatures similar to those of \citet{2015ApJ...814L...2F} between 3\,000 and 8\,000\,K.

The inferred abundance ratio of Ni and Fe from \citet{2018MNRAS.477.3567M} and this work deviate by about 1.5~$\sigma$ for the same objects. The differences are mainly due to the placement of the (pseudo-)continuum across the 7\,200\,\AA\,region, leading to a different line ratio of \ion{Fe}{II} 7\,155\,\AA\,and \ion{Ni}{II} 7\,378\,\AA. In this work we opted for a conservative continuum placement as most of it can be explained by a blend of weak lines of other singly and doubly ionized iron group ions (e.g. [\ion{Co}{III}], [\ion{Fe}{III}]). The departure coefficients corresponding to the allowed range of temperatures and densities (see Fig.~\ref{FigureTemperatureDensity}) of the emitting material are in good agreement with the ones used by \citet{2018MNRAS.477.3567M}. The use of the \ion{Fe}{II} NIR/VIS relation allows us to better constrain the allowed range of the physical parameters of the singly ionized ejecta, leading to reduced uncertainties compared to \citet{2018MNRAS.477.3567M}. We want to emphasise that both works make use of the same atomic data for the ions in question.
\begin{figure}
	\includegraphics[width=\linewidth]{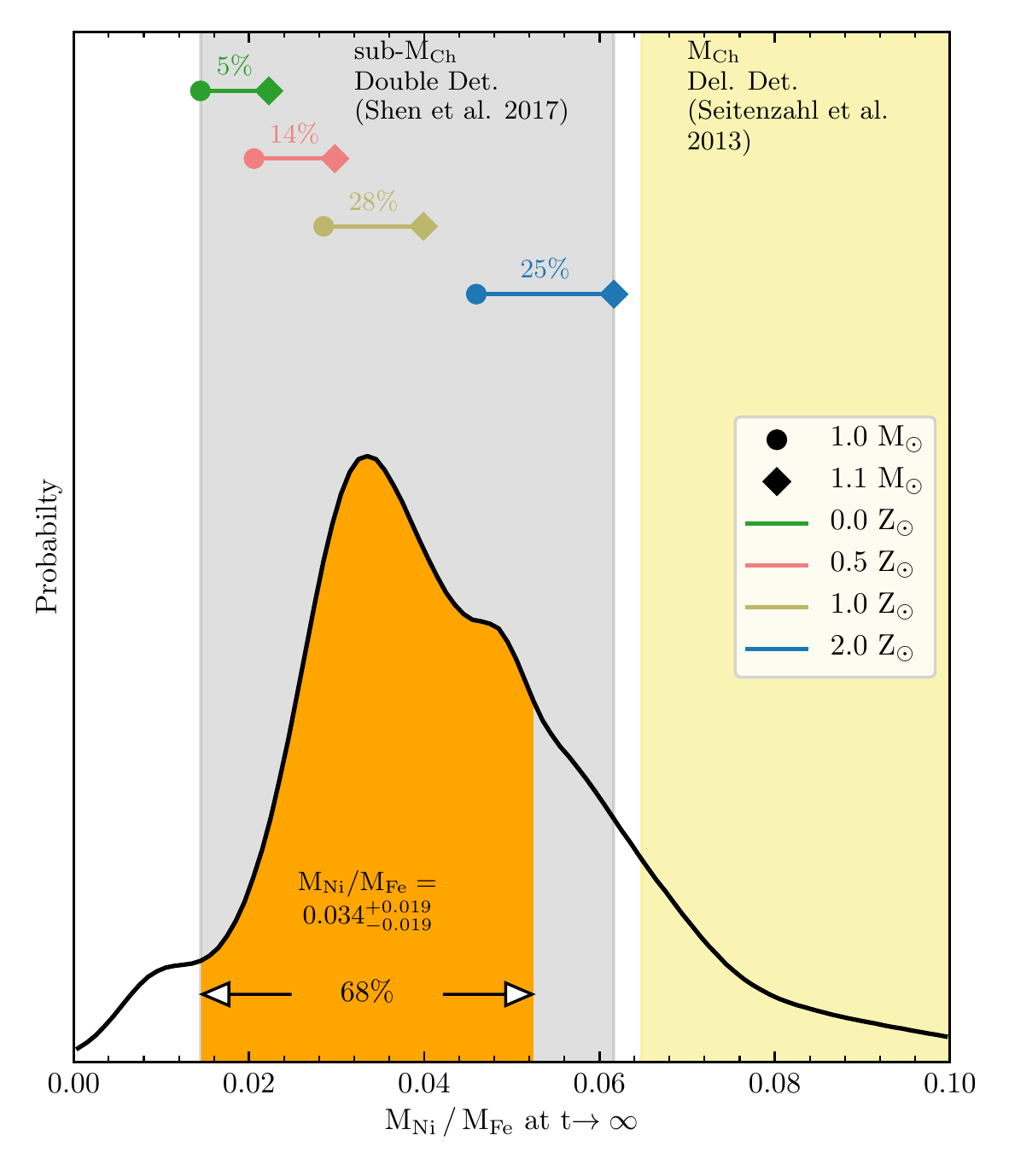}
    \caption{The distribution of the Ni/Fe ratio at $t\rightarrow\infty$ from all available nebular phase spectra (see Table~\ref{TableSpectraOverview}). The Ni/Fe ratio from only optical spectra was corrected according to Section~\ref{TestNIRVIS_relation} by $\sigma_{\mathrm{sys}}=-0.0033^{+0.0037}_{-0.0041}$. The results for SNe with multiple observations were combined so that every supernova in the sample contributes equally to the shown distribution - irrespective of the number of spectra. For each unique SN we drew 100\,000 samples from the posterior distribution of the Ni/Fe mass ratio. The orange shaded region indicates the region containing 68\% of the posterior probability density. The shaded bands display predictions of the Ni to Fe mass ratio from sub-M$_{\text{Ch}}$ \citep[][left]{2018ApJ...854...52S} and M$_{\text{Ch}}$ \citep[][right]{2013MNRAS.429.1156S} explosion model simulations. For sub-M$_{\text{Ch}}$ explosions we also show the range of models for 4 progenitor metallicities and their enclosed fraction of the posterior distribution of our sample.} 
    \label{NiFe_Distribution}
\end{figure}
\subsection{Implications on the explosion mechanism}

\label{Implications}The various theoretical explosion models of SNe Ia predict different amounts of neutron rich material. In M$_{\text{Ch}}$ explosions the amount of synthesized neutron-rich material is determined by two processes: \textit{Carbon simmering} and \textit{neutron-rich burning}:
 
\textit{Carbon simmering} occurs when a white dwarf accretes slowly towards the M$_{\text{Ch}}$. Densities and temperatures in the center become high enough to ignite carbon, but no thermonuclear runaway happens due to a large convective core that allows for cooling through escaping neutrinos \citep{2004ApJ...607..921W, 2004ApJ...616.1102W, 2008ApJ...678.1158P}. The burning of carbon leads to mostly $^{13}$N and $^{23}$Na, which can subsequently capture electrons which further increases the neutron excess \citep{2008ApJ...677..160C, 2016ApJ...825...57M}.

\textit{Neutron-rich burning} to NSE can shift the equilibrium away from $^{56}$Ni to more neutron-rich isotopes ($^{54,56}$Fe, $^{57,58}$Ni, $^{55}$Mn) \citep{1999ApJS..125..439I, 2000ApJ...536..934B}. Just before the explosion, the high central density of the progenitor white dwarf leads to neutronization through electron capture in the densest region. Neutron-rich NSE burning is only possible if there is a neutron excess in the NSE burning central region. 

In sub-M$_{\text{Ch}}$ models such processes are not possible as their progenitors cannot reach the required central density. However, an overabundance of neutrons in a high metallicity progenitor can still lead to the production of neutron-rich IGE \citep{2003ApJ...590L..83T}. The fraction of neutron rich to normal material can cover a wide range of values - from close to zero for $Z=Z_\odot$ to that of M$_{\text{Ch}}$ explosions at several times solar metallicity \citep{2018ApJ...854...52S}. It remains to be seen whether such extremely-high metallicity progenitors really exist.

We focus on the neutron-rich, stable $^{58}$Ni. The presence of a signature line close to $7\,378$\,\AA\,reveals that at least some amount of $^{58}$Ni can be found in all normal SNe Ia observed so far. As shown in Fig.~\ref{SN2015F_NiII} the 7\,200\,\AA\,feature can be explained by a blend of mainly [\ion{Fe}{II}] and [\ion{Ni}{II}]. In principle there will also be varying amounts of stable iron produced during the explosion, but this contribution to the total iron mass is hard to disentangle from the overwhelming fraction of daughter products of radioactive $^{56}$Co.

In contrast to the artificial W7 model \citep{1984ApJ...286..644N}, state-of-the-art explosion simulations from both the sub-M$_{\text{Ch}}$ and M$_{\text{Ch}}$ channels show that $^{58}$Ni and $^{56}$Ni are not produced in geometric isolation. The forbidden emission lines of \ion{Fe}{II} and \ion{Ni}{II} in nebular spectra of normal SNe\,Ia exhibit similar widths and shifts, pointing towards a shared emission region. If indeed $^{58}$Ni and $^{56}$Ni share the volume and excitation conditions then the derived mass ratio of \ion{Fe}{II} and \ion{Ni}{II} should be representative for Fe/Ni produced in the explosion.

The observed spectra are fit well with our emission model. By using the relation from Section~\ref{Calibration} we can compute the Ni/Fe ratio. At early times the ratio is still evolving with time as not all the $^{56}$Co has decayed to $^{56}$Fe yet. At late times (>250 days) the ratio remains constant. We find a large spread of Ni/Fe ratios, ranging from 0.02 to 0.08 within the 95\% confidence interval. We do not find any objects for which we can exclude the contribution of Ni to the nebular phase spectrum. 

Our results are in good agreement with sub-M$_{\text{Ch}}$ explosions of solar- to super-solar metallicity progenitors. Only few objects have a Ni to Fe ratio that is consistent with explosion predictions from zero-metallicity sub-M$_{\text{Ch}}$ white dwarfs. There are only few calculations of non-zero metallicity sub-M$_{\text{Ch}}$ explosions \citep{2010ApJ...714L..52S, 2018ApJ...854...52S}. Our data are consistent with both sub-M$_{\text{Ch}}$ detonations and double detonations, but they do not allow us to distinguish between these two scenarios. 

We find a few objects which have Ni/Fe abundances consistent with nucleosynthetic predictions of exploding M$_{\text{Ch}}$ white dwarfs. However, we do not find separate populations but instead the distribution displays a tail of objects which have high Ni/Fe abundances. The abundance distribution of objects which have nebular phase observations peaks at M$_{\text{Ni}}$\,/\,M$_{\text{Fe}}$ = 0.034 with an $68\%$ confidence region between 0.015 and 0.053. $85\,\%$ of the total probability density falls within the shaded band of sub-M$_{\text{Ch}}$ explosion predictions. Our resulting distribution of the Ni/Fe abundance agrees well with the results of \citet{2019arXiv190610126K}, who determined the Ni/Fe abundance from stellar populations of dwarf galaxies. Only $11\,\%$ of the total probability lies in the range of M$_{\text{Ch}}$ delayed-detonation predictions. The presence of both channels is in agreement with findings from nearby SN remnants \citep{2019arXiv190605972S}.

For sub-M$_{\text{Ch}}$ we can compare our resulting distribution to explosion yields of progenitors with different masses and metallicities. Progenitors with masses of 0.9\,M$_\odot$ or less do not produce enough $^{56}$Ni ($<\,0.3\,$M$_\odot$) to explain the brightness of normal SN Ia and are thus discarded for this comparison. The overlap between the range of yields from $1.0\,$M$_\odot$ to $1.1\,$M$_\odot$ progenitors with our inferred Ni/Fe distribution is shown in Fig.~\ref{NiFe_Distribution}. We find good agreement with progenitors between 0.5 and 2 Z$_\odot$. 

\section{Conclusions}
 The 7\,200\,\AA\,feature in nebular spectra of SNe Ia is composed of emission from \ion{Fe}{II} and \ion{Ni}{II} and is present in all objects for which this wavelength region has been observed. The relative contributions of the two ions to the feature vary between different SNe. We have presented a method that allows us to place prior constraints on the N$_e$ and T and applied it to more than 100 optical archival spectra allowing us to determine the distribution of the Ni/Fe ratio for all objects in our sample. 
Our main results are:
\begin{itemize}
    \item[i)] The \ion{Fe}{II} emission in the nebular phase can be described by purely thermal forbidden line emission, and it is in agreement with an expanding and cooling nebula. 
    \item[ii)] The strongest [\ion{Fe}{II}] lines in the NIR and at optical wavelengths evolve with time, and the evolution seems to be very homogeneous across our sample. We obtained a relation that describes the evolution of this line ratio. The ratio does not depend on the atomic data. The evolution of the \ion{Fe}{II} lines can be used to test more sophisticated spectral synthesis calculations of explosion model predictions -- spectra that have been computed from explosion models need to be able to reproduce this relation.
    \item[iii)] The 7\,200\,\AA\,feature only contains \ion{Fe}{II} and \ion{Ni}{II} in normal SNe Ia. A contribution of \ion{Ca}{II}] to this feature would have to be very limited in strength. We used the 19\,390\,\AA\,line to constrain the 7\,378\,\AA\,line for SN\,2015F and SN\,2017bzc as these two lines originate from the same upper level. We find no evidence that \ion{Ca}{II}] emission is required to reproduce the 7\,200\,\AA\,feature.
    \item[iv)] For all objects in the extended sample of more than 100 nebular phase spectra we find that the lines of singly ionized \ion{Fe}{II} and \ion{Ni}{II} have similar widths and shifts and thus come from the same emitting region and share the same physical conditions. For objects for which NIR spectra are available we can extend this claim to \ion{Co}{II} as well. 
    \item[v)] The display of 130 nebular phase spectra shows a large variety in the relative strengths of the \ion{Fe}{II} and \ion{Ni}{II} lines in the 7\,200\,\AA\,feature. Translating the relative line strengths into a mass ratio of the singly ionized species results in a distribution which is expected from mainly sub-M$_{\text{Ch}}$ explosions.
    \item[vi)] We do not find separate populations of sub-M$_{\text{Ch}}$ and M$_{\text{Ch}}$ explosions. However, the high abundance tail of the distribution extends into the M$_{\text{Ch}}$ regime. 11\% of the total probability distribution lies within the M$_{\text{Ch}}$ predictions of the Ni/Fe abundance.
\end{itemize}
\section*{Acknowledgements}
AF thanks Christian Vogl, Kate Maguire, Luke Shingles and Stuart Sim for stimulating discussions during various stages of this project. The authors thank the anonymous reviewer for valuable comments. We thank the staff at the Paranal observatory. This research would not be possible without their efforts in supporting service mode observing. This research has made use of the NASA/IPAC Extragalactic Database (NED) which is operated by the Jet Propulsion Laboratory, California Institute of Technology, under contract with the National Aeronautics and Space Administration. This work made use of the Heidelberg Supernova Model Archive (HESMA), https://hesma.h-its.org. AF acknowledges the support of an ESO Studentship. 

This research made use of Astropy, a community-developed core Python package for Astronomy \citep{2013A&A...558A..33A, 2018AJ....156..123A}, as well as numpy \citep{Walt:2011:NAS:1957373.1957466}, scipy \citep{scipy}, pandas \citep{mckinney-proc-scipy-2010} and matplotlib \citep{doi:10.1109/MCSE.2007.55}.



\bibliographystyle{mnras}
\bibliography{bibliography} 




\appendix
\section{VLT nebular spectra}
\begin{figure}
	\includegraphics[width=\linewidth]{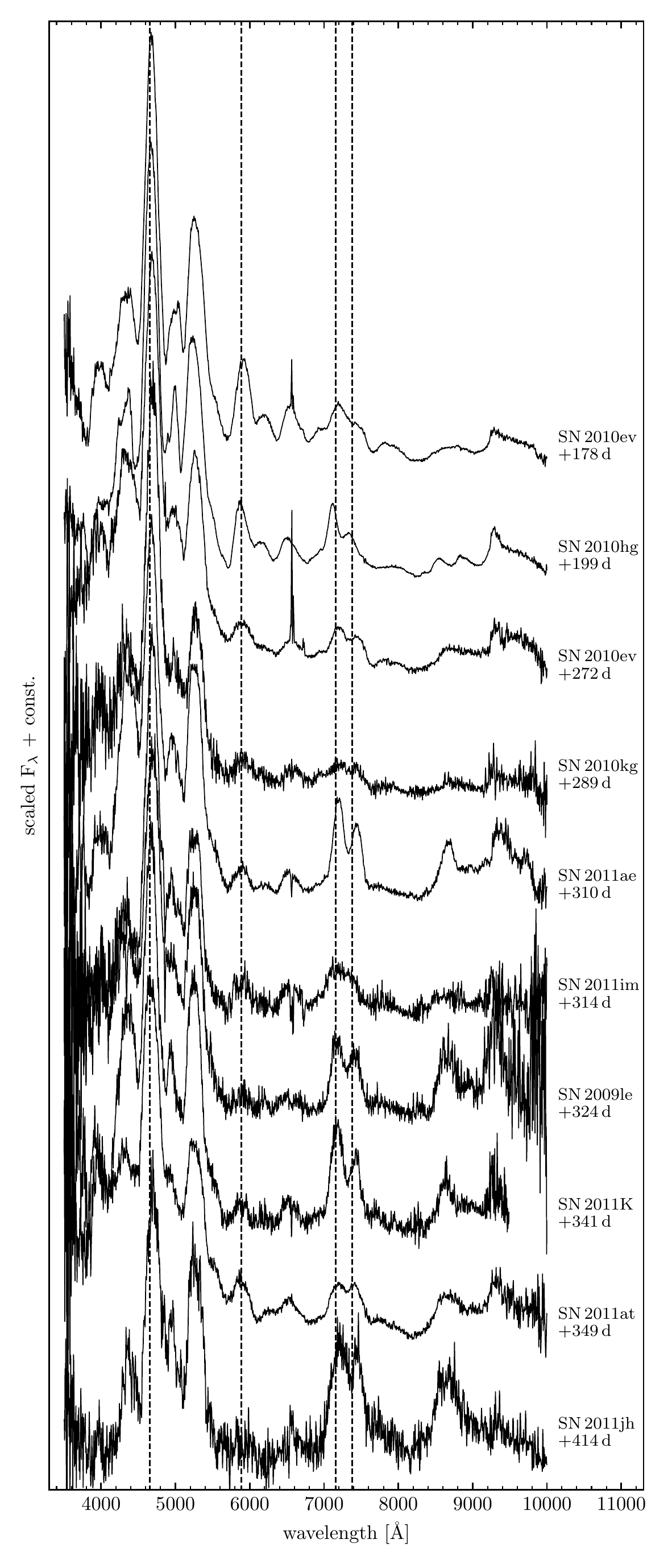}
    \caption{Spectra obtained at the VLT with FORS2 (PI: S. Taubenberger, programme ids: 086.D-0747, 087.D-0161, 088.D-0184, 090.D-0045). The rest wavelengths of the 4\,659\,\AA\,[\ion{Fe}{III}], the 5\,888\,\AA\,[\ion{Co}{III}], the 7\,155\,\AA\,[\ion{Fe}{II}] and the 7\,378\,\AA\,[\ion{Ni}{II}] lines are indicated as dashed lines.}
    \label{SpectraVLT}
\end{figure}

Fig.~\ref{SpectraVLT} presents previously unpublished spectra obtained at the VLT with the FORS2 spectrograph (PI: S. Taubenberger, programme ids: 086.D-0747, 087.D-0161, 088.D-0184, 090.D-0045). The spectra have been corrected for redshift and galactic extinction to better illustrate the position of the strongest Iron, Nickel and Cobalt lines (dashed vertical lines). Additional information on these observations can be found in Table~\ref{TableSpectraOverview}.
\section{Overview of nebular spectra }
In Table~\ref{TableSpectraOverview}, we provide the SN name, subtype, combined galactic and host galaxy color excess, redshift and the date of B-band maximum for each SN Ia that is used in the analysis. Multiple observations of the same SN Ia are sorted by increasing epoch. We also show the telescope and instrument that was used to obtain the spectrum. The measured Ni/Fe mass ratio in the limit $t\rightarrow\infty$ is also given in Table~\ref{TableSpectraOverview} for each spectrum. 
\begin{table*} 
    \caption{Overview of spectra observations.}
    \label{TableSpectraOverview}	
    \centering 
    \scalebox{0.95}{
	\begin{tabular}{l c c cccccccc} 
    \hline
    \hline 
    \centering 
    Supernova & Subtype & E$(B-V)$ & z & Date of max.& Epoch & Telescope & Instrument &  Ref & Ref & M$_{\text{Ni}}$\,/\,M$_{\text{Fe}}$\\
              &         & (mag)    &   &             &       &           &            & Spec & Ext &  ($t\rightarrow\infty$) \\
    \hline  
    SN\,1990N  & Ia-norm   & $0.0223$ & 0.003395 & 10 July 1990 & +186   & WHT-4.2m & FOS-2           & 1 &-& 0.027$^{+0.011}_{-0.010}$\\
             &           &          &          &              & +227   & WHT-4.2m & FOS-2           & 1 & & 0.023$^{+0.006}_{-0.004}$ \\
             &           &          &          &              & +255   & WHT-4.2m & FOS-2           & 1 & & 0.023$^{+0.005}_{-0.004}$ \\
             &           &          &          &              & +280   & WHT-4.2m & FOS-2           & 1 & & 0.023$^{+0.005}_{-0.004}$ \\
             &           &          &          &              & +333   & WHT-4.2m & FOS-2           & 1 & & 0.026$^{+0.007}_{-0.004}$ \\
    SN\,1991T  & 91T-like  & $0.16$   & 0.005777 & 28 Apr 1991  & +258   & WHT-4.2m & ISIS            & 1 & 2& 0.032$^{+0.007}_{-0.005}$ \\
             &           &          &          &              & +316   & INT-2.5m & FOS             & 1 & & 0.031$^{+0.011}_{-0.009}$ \\
             &           &          &          &              & +320.4 & Lick-3m & KAST             & 3 & & 0.033$^{+0.006}_{-0.006}$ \\
             &           &          &          &              & +349.4 & Lick-3m & KAST             & 3 & & 0.027$^{+0.011}_{-0.008}$ \\
    SN\,1993Z  & Ia-norm   & $0.0370$ & 0.004503 & 28 Aug 1993 & +201   & Lick-3m & KAST             & 3 &-& 0.040$^{+0.010}_{-0.007}$ \\
             &           &          &          &              & +233   & Lick-3m & KAST             & 3 & & 0.033$^{+0.007}_{-0.006}$ \\
    SN\,1994ae & Ia-norm   & $0.096$  & 0.004266 & 29 Nov 1994  & +368   & MMT & MMT-Blue             & 4 & 5 & 0.050$^{+0.009}_{-0.008}$ \\
    SN\,1995D  & Ia-norm   & $0.0484$ & 0.006561 & 20 Feb 1995  & +276.8 & MMT & MMT-Blue             & 4 & - & 0.006$^{+0.004}_{-0.004}$  \\
             &           &          &          &              & +284.7 & MMT & MMT-Blue             & 4 & & 0.008$^{+0.006}_{-0.006}$ \\
    SN\,1996X  & Ia-norm   & $0.0596$ & 0.008876 & 18 Apr 1996  & +246   & ESO-1.5m & BC-ESO          & 6 & - & 0.053$^{+0.020}_{-0.018}$ \\
    SN\,1998aq & Ia-norm   & $0.0122$ & 0.003699 & 27 Apr 1998  & +211.5 & FLWO-1.5m & FAST           & 7 & - & 0.043$^{+0.012}_{-0.010}$ \\
             &           &          &          &              & +231.5 & FLWO-1.5m & FAST           & 7 & & 0.052$^{+0.012}_{-0.008}$ \\
             &           &          &          &              & +241.5 & FLWO-1.5m & FAST           & 7 & & 0.037$^{+0.011}_{-0.010}$ \\
    SN\,1998bu & Ia-norm   & $0.34$   & 0.002992 & 19 May 1998  & +179.5 & FLWO-1.5m & FAST           & 8&9&0.052$^{+0.005}_{-0.005}$ \\
             &           &          &          &              & +190.5 & FLWO-1.5m & FAST           & 8& & 0.046$^{+0.007}_{-0.006}$ \\
             &           &          &          &              & +208.5 & FLWO-1.5m & FAST           & 8& & 0.047$^{+0.008}_{-0.007}$ \\
             &           &          &          &              & +217.5 & FLWO-1.5m & FAST           & 8& & 0.049$^{+0.007}_{-0.006}$ \\
             &           &          &          &              & +236.4 & Lick-3m & KAST             & 3 & & 0.051$^{+0.007}_{-0.006}$ \\
             &           &          &          &              & +243.5 & FLWO-1.5m & FAST           & 8& & 0.050$^{+0.008}_{-0.007}$ \\
             &           &          &          &              & +249   & Danish-1.54m & DFOSC       & 10& & 0.058$^{+0.009}_{-0.008}$ \\
             &           &          &          &              & +280.4 & Lick-3m & KAST             & 3 & & 0.059$^{+0.008}_{-0.008}$ \\
             &           &          &          &              & +329   & ESO-3.6m & EFOSC2-3.6      & 10& & 0.058$^{+0.010}_{-0.011}$ \\
             &           &          &          &              & +340.3 & Lick-3m & KAST             & 3 & & 0.055$^{+0.015}_{-0.011}$ \\
             &           &          &          &              & +347.3 & VLT & FORS1                & 11& & 0.061$^{+0.010}_{-0.009}$ \\
    SN\,1999aa & 91T-like  & $0.0342$ & 0.014907 & 26 Feb 1999  & +256.6 & Keck1 & LRIS               & 3 & - & 0.053$^{+0.009}_{-0.008}$ \\
             &           &          &          &              & +282.6 & Keck1 & LRIS               & 3 & & 0.055$^{+0.012}_{-0.010}$ \\
    SN\,2002bo & Ia-norm   & $0.53$   & 0.0043   & 24 Mar 2002  & +227.7 & Keck2    & ESI           & 3 & 12 & 0.051$^{+0.010}_{-0.009}$ \\
    SN\,2002cs & Ia-norm   & $0.088$  & 0.015771 & 16 May 2002  & +174.2 & Keck2    & ESI           & 3 &-& 0.059$^{+0.018}_{-0.017}$ \\
    SN\,2002dj & Ia-norm   & $0.096$  & 0.009393 & 24 Jun 2002  & +222   & ESO-NTT  & EFOSC2-NTT    & 13& 13 & 0.046$^{+0.012}_{-0.011}$ \\
             &           &          &          &              & +275   & VLT-UT1  & FORS1         & 13&    & 0.051$^{+0.010}_{-0.008}$ \\
    SN\,2002er & Ia-norm   & $0.36$   & 0.009063 & 06 Sept 2002 & +216   & TNG      & DOLORES       & 14& 15 & 0.083$^{+0.019}_{-0.016}$ \\ 
    SN\,2003cg & Ia-norm   & $1.33$   & 0.004113 & 31 Mar 2003  & +385   & VLT-UT1  & FORS2         & 16& 16 & 0.049$^{+0.008}_{-0.007}$ \\
    SN\,2003du & Ia-norm   & $0.0081$ & 0.006408 & 06 May 2003  & +209   & CA-3.5m  & MOSCA         & 17&-& 0.034$^{+0.005}_{-0.005}$ \\
             &           &          &          &              & +221   & CA-2.2m  & CAFOS         & 17& & 0.039$^{+0.007}_{-0.006}$ \\
             &           &          &          &              & +272   & CA-3.5m  & MOSCA         & 17& & 0.031$^{+0.009}_{-0.008}$ \\
             &           &          &          &              & +377   & TNG      & DOLORES       & 17& & 0.028$^{+0.004}_{-0.004}$ \\
    SN\,2003gs & Ia-norm   & $0.066$  & 0.004770 & 28 July 2003& +201   & Keck2    & ESI           & 3 & 18 & 0.054$^{+0.008}_{-0.007}$ \\
    SN\,2003hv & Ia-norm   & $0.0133$ & 0.005624 & 06 Sept 2003 & +323   & VLT-UT1  & FORS2         & 19&-& 0.087$^{+0.012}_{-0.010}$ \\
    SN\,2003kf & Ia-norm   & $0.269$  & 0.007388 & 11 Dez 2003  & +397.3 & Magellan-Clay & LDSS-2   & 4 & - & 0.032$^{+0.007}_{-0.006}$ \\
    SN\,2004bv & 91T-like  & $0.0546$ & 0.010614 & 17 May 2004 & +161   & Keck1    & LRIS          & 3 &-& 0.058$^{+0.010}_{-0.009}$ \\
    SN\,2004eo & Ia-norm   & $0.093$  & 0.015718 & 30 Sept 2004 & +228   & VLT-UT1  & FORS2         & 20& - & 0.055$^{+0.009}_{-0.008}$ \\
    SN\,2005cf & Ia-norm   & $0.20$   & 0.006461 & 12 Jun 2005  & +267   & Gemini-N & GMOS          & 21& 22& 0.030$^{+0.005}_{-0.005}$ \\
             &           &          &          &              & +319.6 & Keck1    & LRIS          & 22&   & 0.029$^{+0.005}_{-0.006}$  \\
    SN\,2006dd & Ia-norm   & $0.083$   & 0.005871 & 03 July 2006 & +195   & LCO-duPont & WFCCD       & 23& 23 & 0.044$^{+0.006}_{-0.005}$ \\ 
    SN\,2006X  & Ia-norm   & $1.46$    & 0.005294 & 19 Feb 2006  & +277.6 & Keck1    & LRIS	       & 24&12  & 0.065$^{+0.008}_{-0.007}$ \\
             &           &           &          &              & +360.5 & Keck1    & LRIS          & 3 &    & 0.063$^{+0.010}_{-0.009}$ \\
    SN\,2007af & Ia-norm   & $0.181$   & 0.005464 & 16 Mar 2007 & +301   & MMT      & MMT-Blue      & 4 & 12 & 0.035$^{+0.006}_{-0.005}$ \\
    SN\,2007le & Ia-norm   & $0.40$    & 0.006721 & 27 Oct 2007  & +304.7 & Keck1    & LRIS          & 3 & 12 & 0.032$^{+0.005}_{-0.005}$ \\
    SN\,2007sr & Ia-norm   & $0.17$    & 0.005477 & 16 Dez 2007  & +190   & Magellan-Clay & LDSS-3   & 4 & -  & 0.031$^{+0.006}_{-0.005}$ \\
    SN\,2008Q  & Ia-norm   & $0.0716$  & 0.008016 & 09 Feb 2008  & +201.1 & Keck1    & LRIS          & 3 & -  & 0.081$^{+0.011}_{-0.010}$ \\
    SN\,2009ig & Ia-norm   & $0.049$   & 0.008770 & 06 Sept 2009 & +405   & VLT-UT1  & FORS2         & 25& 12 & 0.028$^{+0.005}_{-0.005}$ \\
    SN\,2009le & Ia-norm   & $0.111$   & 0.017786 & 26 Nov 2009  & +324   & VLT-UT1  & FORS2         & TW& 12 & 0.036$^{+0.006}_{-0.005}$ \\
    SN\,2010ev & Ia-norm   & $0.41$    & 0.009211 & 05 July 2010 & +178   & VLT-UT1  & FORS2         & TW& 12 & 0.038$^{+0.008}_{-0.007}$ \\
             &           &           &          &              & +272   & VLT-UT1  & FORS2         & TW&    & 0.044$^{+0.007}_{-0.006}$ \\
    SN\,2010gp & Ia-norm   & $0.21$    & 0.024480 & 25 July 2010 & +279   & VLT-UT1  & FORS2         & 25& 26 & 0.033$^{+0.005}_{-0.005}$ \\
    SN\,2010hg & Ia-norm   & $0.101$   & 0.008219 & 15 Sept 2010 & +199   & VLT-UT1  & FORS2         & TW& -  & 0.033$^{+0.005}_{-0.006}$ \\
    SN\,2010kg & Ia-norm   & $0.130$   & 0.016642 & 11 Dec 2010  & +289   & VLT-UT1  & FORS2         & TW& -  & 0.051$^{+0.012}_{-0.009}$ \\  
        	\hline
    \end{tabular} }
\end{table*}
\begin{table*} 
\contcaption{Overview of spectra observations.}
    \centering 
    \scalebox{0.95}{
	\begin{tabular}{l c c c c c c c c c c} 
    \hline
    \hline 
    \centering 
    Supernova & Subtype & E$(B-V)$ & z & Date of max.& Epoch & Telescope & Instrument &  Ref & Ref & M$_{\text{Ni}}$\,/\,M$_{\text{Fe}}$\\
             &           & (mag)             &          &              &        &          &               &    Spec &  Ext &  ($t\rightarrow\infty$) \\
    \hline  
   
    SN\,2011ae & Ia-norm   & $0.0483$  & 0.006046 & 24 Feb 2011  & +310   & VLT-UT1  & FORS2         & TW& -  & 0.031$^{+0.005}_{-0.005}$ \\  
    SN\,2011at & Ia-norm   & $0.0585$  & 0.006758 & 14 Mar 2011  & +349   & VLT-UT1  & FORS2         & TW& -  & 0.059$^{+0.010}_{-0.009}$ \\
    SN\,2011by & Ia-norm   & $0.0119$  & 0.002843 & 10 May 2011 & +206   & Keck1    & LRIS          & 27 & -  & 0.035$^{+0.006}_{-0.005}$ \\  
             &           &           &          &              & +310   & Keck1    & LRIS          & 27 &    & 0.039$^{+0.006}_{-0.006}$ \\
    SN\,2011ek & Ia-norm   & $0.306$   & 0.005027 & 14 Aug 2011  & +423   & VLT-UT1 & FORS2          & 25& -  & 0.020$^{+0.005}_{-0.006}$ \\
    SN\,2011fe & Ia-norm   & $0.10$    & 0.000804 & 10 Sept 2011 & +174   & WHT-4.2m & ISIS          & 28& 12 & 0.043$^{+0.009}_{-0.007}$ \\
             &           &           &          &              & +205   & Lick-3m  & KAST          & 28&    & 0.042$^{+0.006}_{-0.006}$ \\
             &           &           &          &              & +226   & Lick-3m  & KAST          & 28&    & 0.038$^{+0.006}_{-0.005}$ \\
             &           &           &          &              & +230   & LBT      & MODS1         & 28&    & 0.041$^{+0.006}_{-0.005}$ \\
             &           &           &          &              & +233   & Lijiang-2.4m & YFOSC     & 29&    & 0.036$^{+0.005}_{-0.005}$ \\
             &           &           &          &              & +256   & WHT-4.2m & ISIS          & 30&    & 0.039$^{+0.005}_{-0.005}$ \\
             &           &           &          &              & +259   & WHT-4.2m & ISIS          & 28&    & 0.039$^{+0.005}_{-0.005}$ \\
             &           &           &          &              & +289   & WHT-4.2m & ISIS          & 28&    & 0.040$^{+0.005}_{-0.005}$ \\
             &           &           &          &              & +311   & Lick-3m  & KAST          & 28&    & 0.041$^{+0.006}_{-0.005}$ \\
             &           &           &          &              & +314   & GTC      & OSIRIS        & 31&    & 0.047$^{+0.006}_{-0.006}$ \\
             &           &           &          &              & +347   & WHT-4.2m & ISIS          & 28&    & 0.042$^{+0.006}_{-0.006}$ \\
    SN\,2011im & Ia-norm   & $0.0556$  & 0.016228 & 06 Dec 2011  & +314   & VLT-UT1  & FORS2         & TW& -  & 0.047$^{+0.013}_{-0.011}$ \\
    SN\,2011iv & Ia-norm   & $0.0098$  & 0.006494 & 10 Dec 2011  & +318   & VLT-UT1  & FORS2         & 25& -  & 0.051$^{+0.018}_{-0.015}$ \\
    SN\,2011jh & Ia-norm   & $0.0322$  & 0.007789 & 03 Jan 2012  & +414   & VLT-UT1  & FORS2         & TW& -  & 0.022$^{+0.005}_{-0.006}$ \\
    SN\,2011K  & Ia-norm   & $0.0852$  & 0.014891 & 20 Jan 2012  & +341   & VLT-UT1  & FORS2         & TW& -  & 0.021$^{+0.006}_{-0.005}$ \\
    SN\,2012cg & Ia-norm   & $0.20$    & 0.001458 & 03 Jun 2012  & +279   & Keck1    & LRIS          & 30 & 32& 0.025$^{+0.005}_{-0.005}$ \\
             &           &           &          &              & +339   & VLT-UT2  & XShooter     & 33&  & 0.033$^{+0.006}_{-0.006}$ \\
             &           &           &          &              & +343   & VLT-UT1  & FORS2         & 25&    & 0.029$^{+0.006}_{-0.005}$ \\
    SN\,2012cu & Ia-norm   & $1.02$    & 0.003469 & 27 Jun 2012  & +340   & VLT-UT1  & FORS2         & 25& 30 & 0.044$^{+0.007}_{-0.006}$ \\
    SN\,2012fr & Ia-norm   & $0.0177$ & 0.005457 & 12 Nov 2012  & +222   & ANU-2.3m & WiFeS         & 34&  - & 0.027$^{+0.006}_{-0.005}$ \\
             &           &          &          &              & +261   & ANU-2.3m & WiFeS         & 34&    & 0.030$^{+0.006}_{-0.005}$ \\
             &           &          &          &              & +290   & Gemini-S & GMOS-S        & 35&    & 0.028$^{+0.005}_{-0.005}$ \\
             &           &          &          &              & +340   & SALT     & RSS           & 34&    & 0.028$^{+0.008}_{-0.007}$ \\
             &           &          &          &              & +357   & VLT-UT2  & XShooter     & 33&    & 0.023$^{+0.005}_{-0.005}$ \\
             &           &          &          &              & +367   & ANU-2.3m & WiFeS         & 34&    & 0.027$^{+0.005}_{-0.006}$ \\
             &           &          &          &              & +416   & Gemini-S & GMOS-S        & 35&    & 0.030$^{+0.007}_{-0.006}$ \\
    SN\,2012hr & Ia-norm   & $0.0389$ & 0.007562 & 27 Dec 2012 & +283   & Gemini-S & GMOS-S        & 34& -  & 0.021$^{+0.006}_{-0.006}$ \\
             &           &          &          &              & +368   & ANU-2.3m & WiFeS         & 34&    & 0.018$^{+0.007}_{-0.006}$ \\
    SN\,2012ht & Ia-norm   & $0.0252$ & 0.003556 & 03 Jan 2013  & +433   & VLT-UT2  & XShooter     & 33& -  & 0.009$^{+0.004}_{-0.004}$ \\
    SN\,2013aa & Ia-norm   & $0.1458$ & 0.003999 & 21 Feb 2013  & +187   & SALT     & RSS           & 34& -  & 0.034$^{+0.007}_{-0.007}$ \\
             &           &          &          &              & +204   & ANU-2.3m & WiFeS         & 34&    & 0.033$^{+0.007}_{-0.006}$ \\
             &           &          &          &              & +344   & ANU-2.3m & WiFeS         & 34&    & 0.026$^{+0.005}_{-0.005}$ \\
             &           &          &          &              & +360   & VLT-UT2  & XShooter     & 33&    & 0.030$^{+0.006}_{-0.006}$ \\
             &           &          &          &              & +399   & Gemini-S & GMOS-S        & 35&    & 0.027$^{+0.005}_{-0.006}$ \\
             &           &          &          &              & +425   & VLT-UT2  & XShooter     & 33&    & 0.039$^{+0.006}_{-0.006}$ \\
    SN\,2013cs & Ia-norm   & $0.0788$ & 0.009243 & 26 May 2013  & +261   & Gemini-S & GMOS-S        & 35& -  & 0.027$^{+0.005}_{-0.006}$ \\
             &           &          &          &              & +300   & ANU-2.3m & WiFeS         & 34&    & 0.022$^{+0.010}_{-0.008}$ \\
             &           &          &          &              & +303   & VLT-UT2  & XShooter     & 33&    & 0.029$^{+0.006}_{-0.005}$ \\
    SN\,2013ct & Ia-norm   & $0.0244$ & 0.003843 & 04 Apr 2013  & +229   & VLT-UT2  & XShooter     & 33& -  & 0.029$^{+0.006}_{-0.006}$ \\
    SN\,2013dy & Ia-norm   & $0.338$  & 0.003889 & 28 July 2013 & +160   & Lijiang-2.4m & YFOSC     & 29& 36 & 0.033$^{+0.011}_{-0.009}$ \\
             &           &          &          &              & +179   & Lijiang-2.4m & YFOSC     & 29&    & 0.030$^{+0.007}_{-0.006}$ \\
             &           &          &          &              & +333   & Keck2    & DEIMOS        & 36&    & 0.025$^{+0.005}_{-0.005}$ \\
             &           &          &          &              & +419   & Keck2    & DEIMOS        & 34&    & 0.027$^{+0.004}_{-0.005}$ \\
             &           &          &          &              & +423   & Keck1    & LRIS          & 36&    & 0.028$^{+0.005}_{-0.006}$ \\
             &           &          &          &              & +480   & Keck1    & LRIS          & 36&    & 0.026$^{+0.009}_{-0.009}$ \\
    SN\,2013gy & Ia-norm   & $0.155$   & 0.014023 & 18 Dec 2013  & +276   & Keck2    & DEIMOS        & 34& 37 & 0.053$^{+0.009}_{-0.008}$ \\
             &           &           &          &              & +280   & Keck1    & LRIS          & 35&    & 0.057$^{+0.010}_{-0.008}$ \\
    SN\,2014J  & Ia-norm   & $1.43$    & 0.000677 & 01 Feb 2014  & +212.5 & WHT-4.2m & ACAM          & 38& 39 & 0.034$^{+0.004}_{-0.005}$ \\
             &           &           &          &              & +231   & Keck2    & DEIMOS        & 34&    & 0.029$^{+0.005}_{-0.005}$ \\
             &           &           &          &              & +269   & HCT-2m   & HFOSC         & 40&    & 0.029$^{+0.005}_{-0.004}$ \\
             &           &           &          &              & +282   & ARC 3.5m & DIS           & 30&    & 0.026$^{+0.005}_{-0.006}$ \\
             &           &           &          &              & +351   & HCT-2m   & HFOSC         & 40&    & 0.033$^{+0.007}_{-0.006}$ \\
    ASASSN-14jg& Ia-norm & $0.0128$  & 0.0148   & 31 Oct 2014  & +267   & Gemini-S & GMOS-S        & 35& -  & 0.041$^{+0.007}_{-0.006}$ \\
     &           &           &          &              & +323   & VLT-UT2  & XShooter     & 33&    & 0.039$^{+.008}_{-0.007}$ \\
    ASASSN-15be& Ia-norm   & $0.17$  & 0.0219   & 29 Jan 2015  & +266   & VLT-UT2  & XShooter     & 33& -  & 0.053$^{+0.012}_{-0.013}$ \\
           	\hline
    \end{tabular} }
\end{table*}
\begin{table*} 
\contcaption{Overview of spectra observations.}
    \centering 
    \scalebox{0.95}{
	\begin{tabular}{l c c cccccccc} 
    \hline
    \hline 
    \centering 
    Supernova & Subtype & E$(B-V)$ & z & Date of max.& Epoch & Telescope & Instrument &  Ref & Ref & M$_{\text{Ni}}$\,/\,M$_{\text{Fe}}$\\
             &           & (mag)             &          &              &        &          &               &   Spec &  Ext &  ($t\rightarrow\infty$) \\
    \hline

    SN\,2015F  & Ia-norm   & $0.26$    & 0.00489   & 25 Mar 2015  & +181   & VLT-UT2  & XShooter     & TW & 41& 0.050$^{+0.008}_{-0.007}$ \\
             &           &           &          &              & +225   & VLT-UT2  & XShooter     & TW &   & 0.048$^{+0.006}_{-0.005}$ \\
             &           &           &          &              & +239   & VLT-UT2  & XShooter     & TW &   & 0.045$^{+0.005}_{-0.005}$ \\
             &           &           &          &              & +266   & VLT-UT2  & XShooter     & TW &   & 0.050$^{+0.005}_{-0.004}$ \\
             &           &           &          &              & +280   & Gemini-S & GMOS-S        & 35 &   & 0.052$^{+0.006}_{-0.006}$ \\
             &           &           &          &              & +406   & VLT-UT2  & XShooter     & TW &   & 0.049$^{+0.009}_{-0.009}$ \\
    PSNJ1149 & Ia-norm   & $0.0247$  & 0.005589 & 11 July 2015 & +206   & VLT-UT2  & XShooter     & 33& -  & 0.037$^{+0.008}_{-0.007}$ \\
    SN\,2017bzc& Ia-norm   & $0.0122$  & 0.00536 & 14 Mar 2015  & +215   & VLT-UT2  & XShooter     & TW & - & 0.030$^{+0.005}_{-0.005}$ \\

	\hline
    \end{tabular} }
      \begin{flushleft}
      References:\\
      (1)~\citet{1998AJ....115.1096G}; (2)~\citet{1999AJ....118.1766P}; (3)~\citet{2012MNRAS.425.1789S}; (4)~\citet{2012AJ....143..126B};\\
      (5)~\citet{1996ApJ...467..435W}; (6)~\citet{2001MNRAS.321..254S}; (7)~\citet{2003AJ....126.1489B}; (8)~\citet{2008AJ....135.1598M};\\
      (9)~\citet{1999ApJS..125...73J}; (10)~\citet{2001ApJ...549L.215C}; (11)~\citet{2004A&A...426..547S}; (12)~\citet{2013ApJ...779...38P};\\
      (13)~\citet{2008MNRAS.388..971P}; (14)~\citet{2005A&A...436.1021K}; (15)~\citet{2004MNRAS.355..178P}; (16)~\citet{2006MNRAS.369.1880E};\\
      (17)~\citet{2007A&A...469..645S}; (18)~\citet{2009AJ....138.1584K}; (19)~\citet{2009A&A...505..265L}; (20)~\citet{2007MNRAS.377.1531P};\\
      (21)~\citet{2007AIPC..937..311L}; (22)~\citet{2009ApJ...697..380W}; (23)~\citet{2010AJ....140.2036S}; (24)~\citet{2008ApJ...675..626W};\\
      (25)~\citet{2016MNRAS.457.3254M}; (26)~\citet{2013A&A...554A.127M}; (27)~\citet{2015MNRAS.446.2073G}; (28)~\citet{2015MNRAS.450.2631M};\\
      (29)~\citet{2016ApJ...820...67Z}; (30)~\citet{2015MNRAS.453.3300A}; (31)~\citet{2015MNRAS.448L..48T}; (32)~\citet{2012ApJ...756L...7S};\\
      (33)~\citet{2018MNRAS.477.3567M}; (34)~\citet{2015MNRAS.454.3816C}; (35)~\citet{2017MNRAS.472.3437G}; (36)~\citet{2015MNRAS.452.4307P};\\
      (37)~\citet{2018arXiv180901359H}; (38)~\citet{2016MNRAS.457..525G}; (39)~\citet{2014ApJ...788L..21A}; (40)~\citet{2016MNRAS.457.1000S};\\
      (41)~\citet{2017MNRAS.464.4476C}; TW: This Work
    \end{flushleft}
\end{table*}



\bsp	
\label{lastpage}
\end{document}